\documentclass[acmsmall,screen]{acmart}

\usepackage{xpatch}

\makeatletter
\xpatchcmd{\ps@firstpagestyle}{Manuscript submitted to ACM}{}{\typeout{First patch succeeded}}{\typeout{first patch failed}}
\xpatchcmd{\ps@standardpagestyle}{Manuscript submitted to ACM}{}{\typeout{Second patch succeeded}}{\typeout{Second patch failed}}    \@ACM@manuscriptfalse
\makeatother

\usepackage{comment}
\usepackage{pifont}
\usepackage{pbox}
 \usepackage[flushleft]{threeparttable}

\usepackage{booktabs}

\usepackage{diagbox}
\usepackage{tablefootnote}

\usepackage{amsmath}
\usepackage{caption}

\usepackage{subcaption}

\usepackage{hyperref}
\usepackage{algorithmic}
\usepackage{adjustbox}
\usepackage{multirow}
\usepackage{tabularx}
\usepackage{enumitem}
\usepackage{framed}
\usepackage[linesnumbered,ruled]{algorithm2e}
\SetKwInput{KwInput}{Input} \SetKwInput{KwOutput}{Output} \SetKwInput{KwProcedure}{Procedure}
 \SetCommentSty{mycommfont}
\usepackage[colorinlistoftodos]{todonotes}
\usepackage{graphicx}
\usepackage{textcomp}
\usepackage{xcolor}
\usepackage{multicol}
\usepackage{balance}
\usepackage{subcaption}
\usepackage{epstopdf}
 \AppendGraphicsExtensions{.pdf}
\usepackage{array}
\usepackage{listings}
\definecolor{Gray}{gray}{0.9}
\definecolor{pgreen}{rgb}{0,0.5,0}
\usepackage{amsthm}
\makeatletter
\def\th@plain{%
  \thm@notefont{}
  \itshape 
}
\def\th@definition{%
  \thm@notefont{}
  \normalfont 
} \makeatother
\usepackage{capt-of}

\usepackage{color, colortbl}
\definecolor{grey}{rgb}{0.7,0.7,0.7}

\newcommand{\lstbg}[3][0pt]{{\fboxsep#1\colorbox{#2}{\strut #3}}}
\lstdefinelanguage{diff}{
  basicstyle=\ttfamily\scriptsize,,
  morecomment=[f][\lstbg{red!20}]-,
  morecomment=[f][\lstbg{green!20}]+,
  morecomment=[f][\lstbg{yellow!20}]++,
  morecomment=[f][\textit]{@@},
  texcl=false
}

\usepackage{tcolorbox}
\usepackage{xcolor}
\definecolor{todocolor}{rgb}{0.9,0.1,0.1}
\definecolor{indiagreen}{rgb}{0.07, 0.53, 0.03}
\definecolor{hycolor}{rgb}{0.7,0.7,0.3}
\definecolor{darkbrown}{rgb}{0.4, 0.26, 0.13}

\usepackage{listings}
\usepackage{xcolor}
\definecolor{main-color}{rgb}{0.6627, 0.7176, 0.7764}
\definecolor{string-color}{rgb}{0.3333, 0.5254, 0.345}
\definecolor{key-color}{rgb}{0.8, 0.47, 0.196}

\lstdefinestyle{mystyle} {
    language = Java,
    basicstyle = {\ttfamily \color{main-color}},
    stringstyle = {\color{string-color}},
    keywordstyle = {\color{key-color}},
    keywordstyle = [2]{\color{lime}},
    keywordstyle = [3]{\color{yellow}},
    keywordstyle = [4]{\color{teal}},
    morekeywords = [3]{<<, >>},
    morekeywords = [4]{++},
    basicstyle=\ttfamily\scriptsize,
    commentstyle=\color{blue}\ttfamily,
    morecomment=[f][\lstbg{red!20}]-,
    morecomment=[f][\lstbg{green!20}]+,
    morecomment=[f][\lstbg{yellow!20}]++,
    morecomment=[f][\lstbg{yellow!20}]--,
    morecomment=[f][\textit]{@@},
    breaklines=true,
    texcl=false
}

\lstnewenvironment{bash}[1][]
  {\lstset{language=[latex]TeX}\lstset{%
   numbers=left,numberstyle=\scriptsize,
   commentstyle=\color{pgreen},
   xleftmargin=5pt,
   xrightmargin=5pt,
   aboveskip=\bigskipamount,
   belowskip=\bigskipamount, #1,
}} {}

\lstdefinestyle{testlstcolor}{
    language={sh},
    moredelim=**[is][\color{red}]{~}{~},
    moredelim=**[is][\color{blue}]{<}{>},
    moredelim=**[is][\bfseries]{***}{***},
    moredelim=**[is][\color{green}]{~~}{~~},
    showstringspaces=false,
    basicstyle=\ttfamily,
    literate={\\~}{{\textasciitilde}}1
        {\\<}{{\unichar{"003C}}}1
        {\\>}{{\unichar{"003E}}}1
}

\newcolumntype{L}[1]{>{\raggedright\let\newline\\\arraybackslash\hspace{0pt}}m{#1}}
\usepackage{tikz}

\usepackage{xspace}

\newcommand{\rewrite}[1]{\textcolor{magenta}{Rewrite:{#1}}}

\newcommand{\eclipse}{\textsc{Eclipse}\xspace}
\newcommand{\idea}{\textsc{IntelliJ IDEA}\xspace}
\newcommand{\netbeans}{\textsc{Netbeans}\xspace}
\newcommand{\vscode}{\textsc{Visual Studio Code (VSCode)}\xspace}

\newcommand{\jdt}{\textsc{Java Development Tools (JDT)}\xspace}

\newcommand{\idearepo}{\textsc{JetBrains/intellij-community}\xspace}

\newcommand{\ipdataset}{\textsc{IP}\xspace}
\newcommand{\refactorbench}{\textsc{RefactorBench}\xspace}
\newcommand{\ippdataset}{\textsc{IPP}\xspace}

\newcommand{\makestatic}{\textsc{Make Static}\xspace}

\newcommand{\saferefactor}{\textsc{SAFEREFACTOR}\xspace}
\newcommand{\astgen}{\textsc{ASTGen}\xspace}

\newcommand{\Plus}{\mathord{\begin{tikzpicture}[baseline=0ex, line width=0.5, scale=0.06]\draw (1,0) -- (1,2);
\draw (0,1) -- (2,1);
\end{tikzpicture}}}
\newcommand{\Minus}{\mathord{\begin{tikzpicture}[baseline=0ex, line width=0.5, scale=0.06]
\draw (0,1) -- (2,1);
\end{tikzpicture}}}
\def\HiLir{\leavevmode\rlap{\hbox to \hsize{\color{red!50}\leaders\hrule height .8\baselineskip depth .5ex\hfill}}}
\def\HiLi{\leavevmode\rlap{\hbox to \hsize{\color{blue!50}\leaders\hrule height .8\baselineskip depth .5ex\hfill}}}

\newcolumntype{H}{>{\setbox0=\hbox\bgroup}c<{\egroup}@{}}

\newcolumntype{x}[1]{%
{\centering\hspace{0pt}}p{#1}}%

\definecolor{azure(colorwheel)}{rgb}{0.0, 0.5, 1.0}

\newcommand{\inputProgramCategoryNumber}{eight}
\newcommand{\inputProgramSubcategoryNumber}{38}

\newcommand{\inputProgramJavaLanguageFeatureRatio}{64.4\%}

\newcommand{\inputProgramLambdaExpressionNumber}{39}
\newcommand{\inputProgramLambdaExpressionRatio}{14.4\%}

\newcommand{\inputProgramGenericsNumber}{38}
\newcommand{\inputProgramGenericsRatio}{14.1\%}

\newcommand{\inputProgramEnumNumber}{27}
\newcommand{\inputProgramEnumRatio}{10.0\%}

\newcommand{\inputProgramInnerClassNumber}{19}
\newcommand{\inputProgramInnerClassRatio}{7.0\%}

\newcommand{\inputProgramAnonymousClassNumber}{15}
\newcommand{\inputProgramAnonymousClassRatio}{5.6\%}

\newcommand{\inputProgramAnnotationNumber}{29}
\newcommand{\inputProgramAnnotationRatio}{10.7\%}

\newcommand{\inputProgramCommentRelatedNumber}{18}
\newcommand{\inputProgramCommentRelatedRatio}{6.7\%}

\newcommand{\numberofCollectedBugs}{1651}
\newcommand{\numberofRefactoringBugs}{637}
\newcommand{\numberofBugsWithInputProgram}{518}
\newcommand{\numberofBugsWithInputProgramAndPatch}{326}

\newcommand{\numberofFindings}{12}
\newcommand{\numberofRootCause}{six}
\newcommand{\numberofSymptom}{nine}
\newcommand{\behaviorchange}{Behavior Change}

\newcommand{\foundIssueNumber}{130}
\newcommand{\submittedIssueNumber}{21}
\newcommand{\totalconfirmedIssueNumber}{10}
\newcommand{\submittedIssueFixedNumber}{seven}
\newcommand{\submittedIssueConfirmedNumber}{three}

\newcommand{\totalrefactoringtype}{99}
\newcommand{\coveredrefactoringtype}{30}

\begin{document}

\author{Haibo Wang}
\affiliation{%
 \institution{Concordia University}
 \city{Montreal}
 \state{Quebec}
 \country{Canada}}
 \email{haibo.wang@mail.concordia.ca}

 \author{Zhuolin Xu}
\affiliation{%
 \institution{Concordia University}
 \city{Montreal}
 \state{Quebec}
 \country{Canada}}
 \email{zhuolin.xu@mail.concordia.ca}

 \author{Huaien Zhang}
\affiliation{%
 \institution{Hong Kong Polytechnic University}
 \city{Hong Kong}
 \state{}
 \country{China}}
  \email{cshezhang@comp.polyu.edu.hk}

  \author{Nikolaos Tsantalis}
\affiliation{%
 \institution{Concordia University}
 \city{Montreal}
 \state{Quebec}
 \country{Canada}}
  \email{nikolaos.tsantalis@concordia.ca}

 \author{Shin Hwei Tan}\authornote{Corresponding Author}\affiliation{\institution{Concordia University}\city{Montreal}\country{Canada}}\email{shinhwei.tan@concordia.ca}

\acmConference[Conference 2024]{ACM Conference}{X, 2024}{X, X}

\title{An Empirical Study of Refactoring Engine Bugs}



\begin{abstract}

Refactoring is a critical process in software development, aiming at improving the internal structure of code while preserving its external behavior. Refactoring engines are integral components of modern Integrated Development Environments (IDEs) and can automate or semi-automate this process to enhance code readability, reduce complexity, and improve the maintainability of software products. Like traditional software systems, refactoring engines can generate incorrect refactored programs, resulting in unexpected behaviors or even crashes. In this paper, we present the first systematic study of refactoring engine bugs by analyzing bugs arising in three popular refactoring engines (i.e., \eclipse, \idea, and \netbeans). We analyzed these bugs according to their refactoring types, symptoms, root causes, and triggering conditions. We obtained \numberofFindings{} findings and provided a series of valuable guidelines for future work on refactoring bug detection and debugging. Furthermore, our transferability study revealed \foundIssueNumber{} new bugs in the latest version of those refactoring engines. Among the \submittedIssueNumber{} bugs we submitted, \totalconfirmedIssueNumber{} bugs are confirmed by their developers, and \submittedIssueFixedNumber{} of them have already been fixed.

\end{abstract}

\maketitle

\section{Introduction}
\label{sec:intro}

Refactoring is defined as the process of changing a software system in such a way that it does not alter the external behavior of the software, yet improves its internal structure ~\cite{becker1999refactoring}. During software development, developers might perform refactoring manually, which is error-prone and time-consuming, or with the help of tools that automate activities related to the refactoring process~\cite{mens2004survey}. Refactoring has been well-studied as an efficient way to improve software quality ~\cite{du2004refactoring,kim2014empirical} as well as an effective way to facilitate software maintenance and evolution ~\cite{kula2018empirical,wahler2016improving}. Refactoring recommendation tools like the built-in refactoring engines in IDEs (e.g., \eclipse~\cite{Eclipse}, \idea~\cite{IDEA}, and \netbeans~\cite{NetBeans}) have been widely used to facilitate software refactoring.


Despite the prevalence of these refactoring automation tools, developers often experience unexpected results due to the flaws in the refactoring engines. Those bugs could silently change the program behavior or induce inconsistencies in the code base, thus producing very serious effects on real-world applications or improving the complexity of maintenance. Past research works have tried to find bugs in these engines~\cite{daniel2007automated,soares2009saferefactor,mongiovi2016scaling,soares2010making,soares2012automated,soares2011identifying,soares2009generating,gligoric2013systematic}; however, effectively finding bugs in refactoring engines remains a challenge due to the diversity of refactoring types, complex refactoring implementations, and lacking a general and in-depth understanding of bugs in the refactoring engines. Although these techniques have been demonstrated to be able to detect some refactoring engine bugs, they still suffer from low effectiveness, specifically, they spend significant resources and time in generating and executing test programs with a relatively low bug-revealing capability. 
Besides, there still lack a general and in-depth understanding of refactoring engine bugs. For example, we are unclear how refactoring engine bugs are induced (i.e., root causes), how these bugs affect the software (i.e., bug symptoms), and how these bugs can be found (i.e., test oracles). To fill those gaps, this paper conducts the first systematic empirical study to analyze bugs in refactoring engines. Particularly, we investigate the following research questions:


\begin{description}[leftmargin=*]
\item[RQ1: (Refactoring Types)] \textbf{What kind of refactorings are more likely to trigger refactoring engine bugs?}

By identifying which refactorings tend to trigger bugs in refactoring engines, researchers could improve the robustness and reliability of refactoring engines by designing more effective automated testing tools focusing more on the bug-prone refactoring types. Refactoring engine developers could pay more attention to the implementation and validation of those bug-prone refactorings.

\item[RQ2: (Bug Symptoms)] \textbf{What are the symptoms of these bugs? How do these bugs affect the refactoring and IDEs?}

This symptom could facilitate the understanding of the consequences of refactoring engine bugs, which helps to triage them and assess their impacts. Moreover, these bug symptoms could help in designing refactoring engine testing techniques with effective test oracles.



\item[RQ3: (Root Causes)] \textbf{What are the root causes of these bugs? What is the occurrence frequency of different root causes of refactoring engine bugs?}

The root causes facilitate the understanding of the nature of refactoring engine bugs, which is helpful to detect, localize, and fix bugs. Moreover, it is worthwhile to explore the root causes specific to refactoring engine bugs and also investigate whether the conclusions on common root causes between refactoring engine bugs and traditional software bugs are consistent or not.

\item[RQ4: (Symptoms \& Root Causes)] \textbf{What is the relation between bug symptoms and root causes in refactoring engines?}

Understanding root causes in RQ3 and symptoms in RQ2 is the first step to investigating refactoring engine bugs. Mapping them (i.e., understanding which root cause is more likely to produce a specific bug symptom), helps in automatically detecting these bugs. 

\item[RQ5: (Triggering Conditions)] \textbf{What are the triggering conditions for the bugs in refactoring engines?}

To guide and facilitate automated testing for refactoring engines, it is important to understand the triggering conditions. For example, understanding the characteristics of the input programs triggering the refactoring engine bugs could help researchers design more effective tools by generating more bug-prone input programs.

\end{description}

Our study is based on refactoring engines in three popular IDEs, namely \eclipse, \idea, and \netbeans as experimental subjects. We studied \numberofBugsWithInputProgram{} bugs that were collected and labeled manually according to a systematic process. From our manual analysis of these refactoring engine bugs, we identified \numberofRootCause{} root causes, \numberofSymptom{} bug symptoms, and \inputProgramCategoryNumber{} main input program characteristics including \inputProgramSubcategoryNumber{} sub-categories, and obtained \numberofFindings{} major findings. Based on these findings, we provide a series of guidelines for refactoring engine bug detection and debugging in the future.

\begin{table*}
\centering
\caption{Our key findings and implications}
\label{tab:findings}
\begin{adjustbox}{width=\textwidth}
\begin{tabular}{l|l} 
\toprule
\multicolumn{1}{c|}{\textbf{Finding}}                                                                                                                                                                                                                                                                                                                                                                                                                                                                                                  & \multicolumn{1}{c}{\textbf{Implication}}                                                                                                                                                                                                                                                                                                                                                                        \\ 
\hline\hline
\begin{tabular}[c]{@{}l@{}}\textit{Finding 1}: “Extract” (149/28.76\%), “Inline” (88/16.99\%), and “Move” (80/15.44\%) \\are the top three most error-prone refactoring types. Existing testing tools only \\cover about one-third (30/99) of the refactoring types.\end{tabular}                                                                                                                                                                                                                                                      & \begin{tabular}[c]{@{}l@{}}Refactoring engine developers should focus on improving the reliability\\~of the most error-prone refactoring, improving test coverage, and \\considering more diverse input programs. Current testing tools only \\cover about one-third of the refactoring types, leaving gaps that could \\be targeted for improvement.\end{tabular}                                              \\ 
\hline
\begin{tabular}[c]{@{}l@{}}\textit{Finding 2}: “Compile Error” (242), “Crash” (106), and “Behavior Change” (66) \\are the top three most common symptoms of refactoring engine bugs.\end{tabular}                                                                                                                                                                                                                                                                                                                                      & \begin{tabular}[c]{@{}l@{}}Although some of the common symptoms (compile error, crash) can \\be easily detected via compiler or execution, there remains challenges \\in detecting bugs related to behavior change. Researchers and \\developers should enhance test oracles and focus on effectively \\identifying behavior changes.\end{tabular}                                                              \\ 
\hline
\begin{tabular}[c]{@{}l@{}}\textit{Finding 3}: Apart from the refactored code itself, other aspects such as \\warning messages and refactoring availability could also be error-prone. \\Current refactoring engines have not thoroughly taken into account various \\side effects that may be incurred by the refactoring.\end{tabular}                                                                                                                                                                                               & \begin{tabular}[c]{@{}l@{}}Incorporating side effects, such as refactoring availability or warning \\messages, into bug detection strategies could provide a broader \\perspective on refactoring correctness. It is important to maintain~\\the consistency between the refactored code and other software \\artifacts (e.g., code comments, and breakpoints).\end{tabular}                                    \\ 
\hline
\begin{tabular}[c]{@{}l@{}}\textit{Finding 4}: Our study found that “Incorrect Transformations” (165) is \\the most common root cause for both \eclipse and \idea, \\accounting for 62 and 103 bugs, respectively.\end{tabular}                                                                                                                                                                                                                                                                          & \begin{tabular}[c]{@{}l@{}}Our study highlights the prominence of incorrect transformations as a \\root cause of bugs. Developers should work towards~ diverse input \\programs and leverage data-driven automation to minimize these \\issues. Researchers could investigate large language models (LLMs)\\~to generate diverse input programs.\end{tabular}                                                   \\ 
\hline
\begin{tabular}[c]{@{}l@{}}\textit{Finding 5}: “Incorrect Preconditions Checking” (62) is the second \\most common root cause. Preconditions need to be updated considering \\the introduction of new language features.\end{tabular}                                                                                                                                                                                                                                                                                                  & \begin{tabular}[c]{@{}l@{}}Preconditions need to be updated considering the introduction of new\\~language features. Considering the evolution of programming \\language and automating the inference of preconditions for refactoring \\could be a promising research direction.\end{tabular}                                                                                                                  \\ 
\hline
\begin{tabular}[c]{@{}l@{}}\textit{Finding 6}: “Incorrect Flow Analysis” (51), and “Incorrect Type Resolving” (35) \\are also non-negligible root causes that are related to the\\ ineffectiveness of the underlying analysis techniques.\end{tabular}                                                                                                                                                                                                                                                                                 & \begin{tabular}[c]{@{}l@{}}The frequent issues of incorrect flow analysis and incorrect type \\resolving highlight the need for more robust analysis techniques\\~in refactoring engines. Improving these areas could significantly\\~reduce bugs caused by them.\end{tabular}                                                                                                                                  \\ 
\hline
\begin{tabular}[c]{@{}l@{}}\textit{Finding 7}: “Incorrect Transformations” can induce all kinds of buggy symptoms \\except for “Refactoring Not Available”, which is mostly exhibited by \\“Failed Selection Parsing”. “Incorrect Transformations”, “Incorrect Preconditions\\ Checking”, and “Incorrect Flow Analysis” are the common root causes for the\\ top three symptoms.\end{tabular}                                                                                                                                          & \begin{tabular}[c]{@{}l@{}}Designing high-quality test oracles that consider the relationship \\between the top three symptoms and root causes can help \\detect a wider variety of bugs.\end{tabular}                                                                                                                                                                                                          \\ 
\hline
\begin{tabular}[c]{@{}l@{}}\textit{Finding 8}: Most of the refactoring bugs (97.1\%) could be triggered by the default\\ initial input options of refactoring engines.\end{tabular}                                                                                                                                                                                                                                                                                                                                                    & \begin{tabular}[c]{@{}l@{}}As most refactoring engine bugs could be triggered by the\\ default configuration, researchers of testing techniques can focus on\\ designing test programs instead of testing configuration-related \\refactoring bugs.\end{tabular}                                                                                                                                                \\ 
\hline
\begin{tabular}[c]{@{}l@{}}\textit{Finding 9a}: Refactoring programs involving Java language features \\are more likely to trigger bugs in refactoring engines, \\they take up 64.4\% of our studied bug reports.\end{tabular}                                                                                                                                                                                                                                                                                                         & \multirow{2}{*}{\begin{tabular}[c]{@{}l@{}}More testing efforts should be paid to the input programs with \\language-specific features, considering 64.4\% of our studied \\bug reports are related to it, and its prevalence in the OSS \\repositories. Language features, particularly lambda expressions, \\generics, and Enum~contribute to most refactoring bugs.\end{tabular}}                            \\ 
\cline{1-1}
\begin{tabular}[c]{@{}l@{}}\textit{Finding 9b}: Lambda expression (39/14.4\%), Java generics (38/14.1\%), \\and Enum (27/10.0\%) are the top three language features triggering \\refactoring engine bugs due to the complexity of type inference, \\limited flow analysis, and complicated usage scenario.\end{tabular}                                                                                                                                                                                                               &                                                                                                                                                                                                                                                                                                                                                                                                                 \\ 
\hline
\begin{tabular}[c]{@{}l@{}}\textit{Finding 10}: Input programs having complex class relationships \\are more likely to result in refactoring engine failure. Among these, \\refactoring involving inner class (19/7.0\%) and anonymous class (15/5.6\%) \\are the top two most bug-prone.\end{tabular}                                                                                                                                                                                                                                 & \begin{tabular}[c]{@{}l@{}}Complex class structures, such as inner and anonymous classes,\\ are prone to triggering bugs. When designing the templates or\\ rules to generate input programs, or mining input programs for\\ testing, researchers should focus on these language features.\end{tabular}                                                                                                         \\ 
\hline
\begin{tabular}[c]{@{}l@{}}\textit{Finding 11}: Annotation-induced (29/10.7\%) refactoring engine \\bugs are the third most in our studied bug reports with input program. \\The reasons include: (1) Lack of consideration of the semantics of annotations, \\(2) Complexity of annotation processing, which may be processed\\ at various phrases: compile-time (e.g., @Override), runtime (e.g., \\@Deprecated), and deployment (e.g. @WebServlet in Java Servlet API),\\ (3) Complex transformations for annotations.\end{tabular} & \begin{tabular}[c]{@{}l@{}}Annotations add complexity due to their semantic meaning and \\wide-ranging effects. Refactoring engines should be enhanced to \\handle these constructs more effectively, particularly by improving \\transformations that respect class and annotation intricacies. \\Future research could explore how refactoring engines can be \\improved to handle them better.\end{tabular}  \\ 
\hline
\begin{tabular}[c]{@{}l@{}}\textit{Finding 12}: Refactoring involving comments interleaved with code might \\cause refactoring to fail (18/6.7\% of our studied bugs). The reasons include:\\(1) Comment-code dependency, and (2) Semantic gaps between code and \\ comments, indicating the needs for comment-aware auto-refactoring tools.\end{tabular}                                                                                                                                                                              & \begin{tabular}[c]{@{}l@{}}There is a gap in research regarding comment-aware\\ refactoring. Researchers should investigate~new methodologies\\ for creating tools that can bridge the semantic gap between code\\ and comments, enabling more robust refactoring that considers \\code-comment dependencies.\end{tabular}                                                                                      \\
\bottomrule
\end{tabular}
\end{adjustbox}
\end{table*}

In summary, this paper has made the following contributions:
\begin{itemize}[leftmargin=*,labelindent=7pt,nosep]
    \item To the best of our knowledge, we conduct the first systematic study to investigate refactoring engine bugs for three popular refactoring tools (i.e., \eclipse, \idea, and \netbeans). We study the refactoring engine bugs from different perspectives (error-prone refactoring types, root causes, bug symptoms, and triggering conditions), and distill \numberofFindings{} findings. Table~\ref{tab:findings} concludes the key findings of our study. The findings of our study aims to encourage better understanding of the refactoring engine bugs, which can potentially benefit developers of refactoring engines and researchers of refactoring engines bugs detection.
    \item From our mined dataset of bug reports from refactoring engines, we curated a dataset of \numberofBugsWithInputProgram{} bugs, namely \refactorbench. The dataset contains categorized bug reports submitted by users of refactoring engines, which serve as the basis of our study as well as future research in this direction. To facilitate further studies in this area, we open-sourced our data~\cite{ourrepourl}.
     \item We conduct a transferability study based on historical bug reports and our findings. As a result, we have found \foundIssueNumber{} new bugs in the latest version of refactoring tools. Among the \submittedIssueNumber{} bugs we submitted, \totalconfirmedIssueNumber{} bugs are confirmed by their developers, and \submittedIssueFixedNumber{} of them have already been fixed.
\end{itemize}

\section{Background}
\label{sec:related}

\subsection{Refactoring Engine}

Refactoring engines are integral components of Integrated Development Environments (e.g., \eclipse, \idea, and \netbeans), which provide automated support to help developers restructure and optimize their code. Even though they differ in the details and integration with each IDE, their overall workflows share similarities~\cite{petito2007eclipse,jemerov2008implementing,garcia2007testing}. Figure~\ref{refactoringworkflow} shows the general workflow of refactoring engines. The input includes the input program and configuration for a refactoring, the output is the refactored program. We explain each step in the refactoring engine as follows:

\begin{figure}[h]
    \centering
    \vspace{-3pt}    \includegraphics[width=0.8\textwidth]{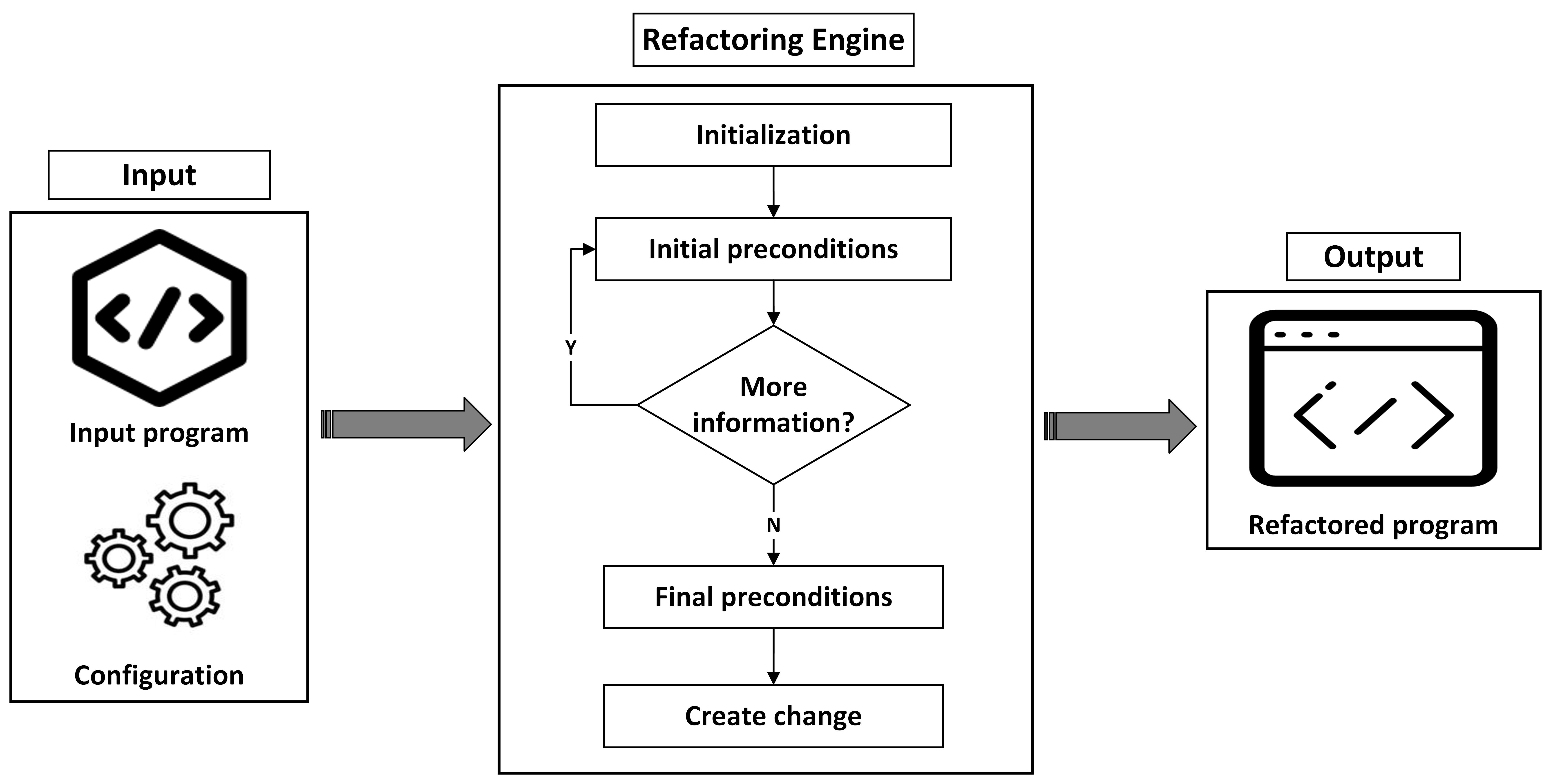}
    \vspace{-6pt}
     \caption{The general workflow of a refactoring engine}
    \label{refactoringworkflow}
\end{figure}

\noindent \textbf{1. Initialization.} In this stage, the refactoring engine initializes the refactoring operation, sets up the necessary context, and identifies the elements to be refactored.



\noindent \textbf{2. Initial conditions checking.} This phase is responsible for checking whether the prerequisites for initiating the refactoring operation are met and checking conditions such as the validity of the selection, potential conflicts, etc. If the initial conditions check fails, the refactoring operation is prevented from proceeding further, and error messages or warnings are displayed to the developers. For example, in the initial conditions checking stage of \makestatic{} refactoring, the method to be refactored should not be a Constructor method~\cite{makeStaticRefactoring}.



\noindent \textbf{3. Additional information.} In this phase, the refactoring engine collects any additional developer inputs or option configurations required for the refactoring. For some refactoring, additional information might be required to perform the refactoring (e.g., the new method name when performing ``Extract Method'' refactoring).


\noindent \textbf{4. Final conditions checking.} After the initial conditions checking has been met and collecting all necessary information to perform the refactoring, the check final conditions can be started to check the remaining preconditions. It examines the potential changes resulting from the refactoring (e.g., potential conflicts with existing code) and verifies that the refactoring can be carried out successfully without bad effects (e.g., behavior-preserving, or quality preconditions~\cite{tsantalis2009identification}). If checking fails, the developer will get a warning or error message. This phase focuses on analysis, verification, and error detection to ensure the safety and feasibility of the refactoring operation. For example, the \makestatic{} refactoring cannot be executed if the refactored method is overridden in a child class because a static method cannot be overridden~\cite{makeStaticRefactoring}.


\noindent \textbf{5. Create change.} This phase is responsible for generating the actual code modifications. The refactoring engine needs to compute the specific changes needed to apply the refactoring, create some change objects representing modifications to be applied, and transform the code. For example, when performing \makestatic{} refactoring for the program, it needs to (1) modify the method declaration (2) update method invocations, and (3) handle related changes (e.g., update Java doc for the refactored method.).

Steps 3 to 5 can be executed repeatedly (e.g., when the developer goes back from the preview page.).

\subsection{Motivation Example}

We will introduce one bug related to \makestatic{} refactoring as an illustrative example to explain: (1) why existing refactoring engine testing tools fail to trigger the bug, and (2) the insights we could get by analyzing existing bug reports.

\noindent \textbf{Eclipse Bug 1083~\cite{makeStaticRefactoringIssue1083}.} Running the initial input program in Figure \ref{lst:make_static_change_method_call} would produce ``Counter: 15''. However, after performing \makestatic{} refactoring  (RQ1: refactoring type is \makestatic{} refactoring) using the refactoring engine of \eclipse on the method ``toBeRefactored()'', the output becomes ``Counter: 5''. The refactored program does not contain any syntax error and \eclipse just silently changes the program behavior without any warning (RQ2: the symptom is ``\behaviorchange''). This happens because \eclipse erroneously interprets a method call within an anonymous class inside the refactored method as a call on ``this'' and thus changes the method invocation to be performed on the added input parameter of the method (RQ3). However, the call to the method ``toCall()'' was made on an instance of the anonymous inner class rather than the outer instance passed to the ``toBeRefactored()'' method. This bug could potentially have multiple effects, as the developer stated ``This is (1) semantically incorrect and (2) leads to compile errors if the type of the added parameter does not provide the method''~\cite{makeStaticRefactoringIssue1083}.


\begin{figure}
\caption{\makestatic{} refactoring erroneously changes method call}
\label{lst:make_static_change_method_call}
{\centering
\begin{adjustbox}{width=0.4\textwidth}

\begin{lstlisting}[style=mystyle, escapechar=^]
public class Foo {
    private int counter;
    public Foo(int initialCounter) {
        this.counter = initialCounter;
    }
-   void toBeRefactored() {
-       new Foo(counter + 10) {
+   static void toBeRefactored(final Foo foo) {
+       new Foo(foo.counter + 10) {
            void toImplement() {
-               toCall();
+               foo.toCall();
            }
        }.toImplement();
    }
    void toCall() {
        System.out.println("Counter: " + counter);
    }
    public static void main(String[] args) {
        Foo foo = new Foo(5);
-       foo.toBeRefactored();
+       Foo.toBeRefactored(foo);
    }
}
\end{lstlisting}\par
\end{adjustbox}\par
} 
\end{figure}

The \saferefactor-based tools~\cite{soares2009saferefactor,mongiovi2016scaling,soares2010making,soares2012automated,soares2011identifying,soares2009generating} cannot identify this bug because they can only generate test cases for the common methods before and after the refactoring in the project, which means they cannot exercise the changed methods (e.g., refactored methods). 
\astgen~\cite{daniel2007automated} may fail to detect this bug because it requires pre-defined templates to generate input programs, however, the developers do not have the knowledge about the types of program structure that are more likely to trigger the refactoring engine bug. As a complementary work, our study could provide insights into the templates or anti-patterns design, improving the effectiveness and efficiency. Based on the analysis for the current \eclipse bug report, 
we conducted a transferability study (refer to Section~\ref{sec:crossstudy} for details) by cross-validating this bug in both \idea and \netbeans, and successfully discovered one new bug in the latest version (2024.1.2 by the time of our study) of \idea (RQ5). We have submitted a bug report~\cite{makeStaticRefactoringIDEAIssue354116} to the \idea issue tracker system, it has been confirmed and fixed by the engine developers.



Apart from the bug in Figure~\ref{lst:make_static_change_method_call}, there are still some other bugs~\cite{makeStaticRefactoringIssue1043,makeStaticRefactoringIssue1044,makeStaticRefactoringIssue1045} related to \makestatic{} refactoring. Those bugs could silently change the program behavior, resulting in compile errors, or inducing inconsistencies in the code base, thus producing unexpected behavior on real-world applications or making it difficult to maintain. Analyzing those bug reports could provide us with a deep understanding of the refactoring engine bug triggering conditions, testing Oracle design, and input program generation. This paper makes the first attempt to comprehensively study the refactoring engine bugs, paves the road for related future research, and provides guidelines for engine developers and researchers to test and improve the reliability of the refactoring engines.

\section{Methodology}
\label{sec:studymet}
\subsection{Dataset Collection}
In this study, we use the three most popular refactoring engines as subjects, including (1) \jdt from \eclipse, (2) the Java refactoring component of \idea, and (3) \netbeans refactoring engine. As Java language support for \vscode is built upon \eclipse JDT~\cite{vscodejava}, we choose to focus on \eclipse and exclude \vscode as our study object. There are some categories of bug reports with different purposes, such as feature requests and questions. Hence, we need to identify the bug reports that aim to fix refactoring engine bugs only. Specifically, following existing studies~\cite{sun2016toward,shen2021comprehensive}, we collect bug reports with titles or discussions that contain at least one refactoring engine bug-relevant keyword (i.e., ``refactoring'' and ``refactor''). Our study aims to investigate the characteristics of refactoring engine bugs, so we focus on bugs that are fixed and not duplicated, together with their corresponding patches~\cite{sun2016toward,shen2021comprehensive,zhong2022enriching}. Specifically, we consider a bug is fixed if its ``Resolution'' field is set to ``FIXED'' and the ``Status'' field is set to ``RESOLVED'', ``VERIFIED'' or ``CLOSED'' in the bug repositories of \eclipse and \netbeans in Bugzilla. For issues after a specific date (Jan 2022 for \netbeans, and Apr 2022 for \eclipse JDT), both \eclipse JDT and \netbeans started to migrate their issue trackers to GitHub. To get a complete list of bug reports, we also crawled issues and patches from their GitHub repositories using the GitHub APIs~\cite{githubapis}. For each repository, we search for the closed issues (resolved bug reports) with the same keywords, since these issues have been confirmed by the developers and are more likely to contain fixes. We further searched for issues that contain at least one corresponding commit. The goal is to get a list of related and closed issues, which have been fixed by the developers. To remove the bug reports that exist in both GitHub and Bugzilla, we de-duplicate them based on their bug ID and title. We only collect fixed bug reports because (1) they are considered to be more important by developers, and (2) such reports contain more information (e.g., patches, and discussions among developers) about the involved bugs, which allows us to have a better understanding of these bugs. For \idea, we collect fixed bugs from its issue tracker using the same keywords. Since the patches are not included in the bug reports in the \idea issue tracker~\cite{ideaIssueTracker}, we need to identify the revisions that correspond to these fixed bugs. As \idea developers usually add the bug report ID as a marker in the commit message, we use the GitHub APIs~\cite{githubapis} to search for commits where the commit message contains the bug report ID in the \idearepo~\cite{ideaGithub} repository. For the issues marked as duplicates by the refactoring engine developers, we will not include them in our dataset.

\begin{table}
{\centering
\caption{Statistical Information of Datasets}
\label{tab:dataset}
\begin{adjustbox}{width=0.7\linewidth}
\begin{tabular}{cccccc} 
\toprule
\textbf{Tool} & \textbf{Duration}                       & \textbf{\#CB} & \textbf{\#RB} & \multicolumn{1}{l}{\textbf{\#IP}} & \multicolumn{1}{l}{\textbf{\#IPP}}  \\ 
\hline
\eclipse       & 2005-03-16 \textasciitilde{} 2023-06-12 & 841           & 289           & 240                               & 122                                \\
\idea & 2016-03-23 \textasciitilde{} 2024-03-22 & 601           & 268           & 209                               & 204                                \\
\netbeans      & 2012-04-08 \textasciitilde{} 2024-04-22 & 209           & 80            & 69                                & 0                                  \\ 
\hline
Total         & -                                       & 1651          & 637           & 518                               & 326                                \\
\bottomrule
\end{tabular}
\end{adjustbox}\par
} 
\begin{tablenotes}
\footnotesize
\item{
\textbf{CB} : Collected bug reports; \textbf{RB} : Refactoring engine bug reports; \textbf{IP} : Refactoring engine bug reports containing input program; \textbf{IPP} : Further filter refactoring engine bug reports containing both input program and patch. Note that we derive our RQ1, RQ2, and RQ5 on the \ipdataset, RQ3 and RQ4 on the \ippdataset.
}
\end{tablenotes}
\end{table}

Table~\ref{tab:dataset} presents detailed information about our dataset. 
As it is a bit too time-consuming to manually analyze all bugs, we only collected bugs for a given period based on their fixing time. As shown in the ``Duration'' column of Table~\ref{tab:dataset}, it includes  
18 years for Eclipse, 8 years for IDEA, and 12 years for NetBeans. The ``\#CB'' column presents the total number of collected bug reports for the given periods. 
Then, two authors manually analyzed these reports (refer to Section~\ref{sec:classification} for the process).
In total, we collect \numberofCollectedBugs{} bug reports. Among them, we identified \numberofRefactoringBugs{} bug reports related to refactoring engine bugs. We remove duplicated bug reports by checking if two bug reports share the same refactoring type, symptom, and similar input program, after this step, \numberofBugsWithInputProgram{} unique bug reports containing input program (\refactorbench dataset) are remained. Due to the migration of the bug-tracking systems and the refactoring tool repositories, some patch links are out of date and no longer available. Finally, we collected \numberofBugsWithInputProgramAndPatch{} bug reports containing both the input program and patch (\ippdataset dataset). Compared to \eclipse and \idea, \netbeans tends to contain fewer refactoring engine bug reports mainly because it is not as widely used as the other two~\cite{idesusagestatus}. We derived our RQ1 (Refactoring Types), RQ2 (Bug Symptoms), and RQ5 (Triggering Conditions) based on the \refactorbench dataset, since the input programs in those bug reports are confirmed by the engine developers that they can trigger the bugs, which means they are valuable for our research. As for RQ3 (Root Causes) and RQ4 (Relations between Symptoms and Root Causes), they require not only analyzing the input programs and discussions in the bug reports but also analyzing the patches to determine the root causes, thus we use the \ippdataset dataset.

\subsection{Classification and Labeling Process} 
\label{sec:classification}
We investigate each bug from four aspects: 1) the refactoring type of the bug, 2) the root cause of the bug, 3) the symptom that the bug exhibits, and 4) the characteristics of the input program triggering the bug. To label the root causes of each bug, we first adopted initial taxonomies of
root causes from the existing related works~\cite{shen2021comprehensive,du2021empirical,xiong2023empirical}, and then adapted them to refactoring engine bugs through an open-coding scheme. Specifically, one author first read all reports to determine the root causes of our collected bugs
based on the initial general taxonomy, further adding refactoring engine specifical categories and removing irrelevant categories. Then, two annotators (i.e., two authors of the paper) independently labeled these
bug reports. Following existing approaches~\cite{garcia2020comprehensive,islam2019comprehensive,wang2023compatibility,win2023towards,shen2021comprehensive}, we measured the inter-rater agreement among the annotators via Cohen’s Kappa coefficient. Particularly, the Cohen’s Kappa coefficient was nearly 40\% for the first 10\% of labeling results, and thus we conducted a training session about labeling. After that, two authors
labeled 20\% of bug reports (including the previous 10\%), and Cohen’s Kappa coefficient was computed as
85\%. After further discussion of the disagreements, Cohen’s Kappa coefficient
was always more than 90\% in subsequent labeling attempts (i.e., labeling 20\% \textasciitilde{} 100\% of bug reports with an interval of 10\%). In each labeling attempt, the two authors discuss their disagreements until they reach a consensus. Finally, all bugs were labeled consistently. During the labeling process, we filter out irrelevant bug reports (e.g., feature requests). We also notice that some developers reported the same input programs causing refactoring engine bugs to multiple tools (e.g., both \eclipse and \netbeans). For these reports, we only count them once. We derive the categories of symptoms, refactoring types, and input program characteristics using the same procedures.

\section{RESULTS AND ANALYSIS}


\subsection{RQ1: Refactoring Types}
\label{sec:pattern}




\begin{table}
\centering
\caption{Statistics of bugs found across different refactoring types}
\label{tab:type}
\resizebox{0.8\textwidth}{!}{
\centering
\begin{tabular}{|ll|ll|ll|} 
\hline
Category         & (\# / \%)     & Subcategory                   & (\#) & Refactoring                             & (\#)  \\ 
\hline
Extract          & 149 / 28.76\% & Inheritance Refactoring       & 61   & Pull Up\textsuperscript{A,S,G}                                 & 31    \\
                 &               &                               &      & Push Down\textsuperscript{A,S,G}                               & 12    \\
                 &               &                               &      & Extract Interface\textsuperscript{G}                       & 8     \\
                 &               &                               &      & Extract Class\textsuperscript{S,G}                           & 5     \\
                 &               &                               &      & Extract Superclass\textsuperscript{G}                      & 5     \\ 
\cline{3-6}
                 &               & Extract Variable              & 47   & Extract Variable\textsuperscript{G}                        & 37    \\
                 &               &                               &      & Extract Constant\textsuperscript{G}                        & 10    \\ 
\cline{3-6}
                 &               & Extract Method                & 41   & Extract Method\textsuperscript{S,G}                          & 40    \\
                 &               &                               &      & Extract Delegate                        & 1     \\ 
\hline
Inline           & 88 / 16.99\%  & Inline Method                 & 46   & Inline Method\textsuperscript{G}                           & 46    \\ 
\cline{3-6}
                 &               & Inline Constant And Variable  & 28   & Inline Variable\textsuperscript{G}                         & 25    \\
                 &               &                               &      & Inline Field                            & 2     \\
                 &               &                               &      & Inline Constant\textsuperscript{G}                         & 1     \\ 
\cline{3-6}
                 &               & Inline Class                  & 13   & Inline Class                            & 11    \\
                 &               &                               &      & Inline Interface                        & 2     \\ 
\cline{3-6}
                 &               & Inline Others                 & 1    & Inline Expression                       & 1     \\ 
\hline
Move             & 80 / 15.44\%  & Move Method                   & 43   & Move Method\textsuperscript{S}                            & 42    \\
                 &               &                               &      & Move Instance Method\textsuperscript{G}                    & 1     \\ 
\cline{3-6}
                 &               & Move Class                    & 30   & Move Type To New File\textsuperscript{G}                   & 20    \\
                 &               &                               &      & Move Inner Class To Outer Level\textsuperscript{A}         & 6     \\
                 &               &                               &      & Move Class                              & 4     \\ 
\cline{3-6}
                 &               & Move Constant And Variable    & 7    & Move Field                              & 4     \\
                 &               &                               &      & Move Constant                           & 2     \\
                 &               &                               &      & Move Parameter                          & 1     \\ 
\hline
Rename           & 49 / 9.46\%   & Rename Method                 & 20   & Rename Method\textsuperscript{A,S,G}                           & 20    \\ 
\cline{3-6}
                 &               & Rename Constant And Variable  & 18   & Rename Field\textsuperscript{A,S,G}                            & 9     \\
                 &               &                               &      & Rename Variable\textsuperscript{S,G}                         & 9     \\ 
\cline{3-6}
                 &               & Rename Class                  & 10   & Rename Class\textsuperscript{A,S}                            & 9     \\
                 &               &                               &      & Rename Interface                        & 1     \\ 
\cline{3-6}
                 &               & Rename Other                  & 1    & Rename Enum                             & 1     \\ 
\hline
Change Signature & 40 / 7.72\%   & Change Method Signature       & 38   & Change Method Signature\textsuperscript{A,S,G}                 & 38    \\
\cline{3-6}
                 &               & Change Class Signature        & 2    & Change Class Signature                  & 2     \\ 
\hline
Introduce        & 25 / 4.83\%   & Introduce Variable            & 15   & Introduce Variable                      & 9     \\
                 &               &                               &      & Introduce Field                         & 4     \\
                 &               &                               &      & Introduce Constant                      & 2     \\ 
\cline{3-6}
                 &               & Introduce Method              & 9    & Introduce Factory\textsuperscript{G}                       & 5     \\
                 &               &                               &      & Introduce Indirection\textsuperscript{G}                   & 2     \\
                 &               &                               &      & Introduce Local Extension               & 1     \\
                 &               &                               &      & Introduce Method                        & 1     \\ 
\cline{3-6}
                 &               & Introduce Class               & 1    & Introduce Parameter Object\textsuperscript{G}              & 1     \\ 
\hline
Convert          & 24 / 4.63\%   & Convert Constant And Variable & 10   & Convert Variable                        & 4     \\
                 &               &                               &      & Convert String To Textblock             & 4     \\
                 &               &                               &      & Convert Boolean                         & 1     \\
                 &               &                               &      & Convert To Automic                      & 1     \\ 
\cline{3-6}
                 &               & Convert Class                 & 8    & Convert Anonymous To Inner ClassClass\textsuperscript{G}        & 7     \\
                 &               &                               &      & Convert Class To Record                 & 1     \\ 
\cline{3-6}
                 &               & Convert Method                & 6    & Convert To Switch Expression Expression  & 2     \\
                 &               &                               &      & Convert To Enhanced For Loops         & 2     \\
                 &               &                               &      & Convert Anonymous To Lambda           & 1     \\
                 &               &                               &      & Convert To Instance Method              & 1     \\ 
\hline
Replace          & 10 / 1.93\%   & Replace Class                 & 6    & Replace Constructor With Builder        & 2     \\
                 &               &                               &      & Replace Constructor With Factory        & 2     \\
                 &               &                               &      & Replace Constructor With Factory Method & 2     \\ 
\cline{3-6}
                 &               & Replace Method                & 4    & Replace With Lambda                     & 2     \\
                 &               &                               &      & Replace Anonymous With Lambda           & 1     \\
                 &               &                               &      & Replace Foreach With For                & 1     \\ 
\hline
Other            & 53 / 10.23\%  & Others                        & 53   & Encapsulate Field                       & 7     \\
                 &               &                               &      & Infer Type Arguments\textsuperscript{G}                              & 7     \\
                 &               &                               &      & Make Static                             & 5     \\
                 &               &                               &      & Use Supertype Wherever Possible\textsuperscript{G}         & 5     \\
                 &               &                               &      & Others\textsuperscript{A,S,G}                                  & 29    \\
\hline
\end{tabular}
}
\begin{tablenotes}
\footnotesize
\item{
The superscript `A' indicates the refactoring type is covered by \astgen; `S' indicates that current refactoring is covered by \saferefactor-based tools; `G' means current refactoring is covered by~\cite{gligoric2013systematic}. Note that some covered refactoring types are merged into the ``Others'' row in the ``Refactoring'' column, so if there is one refactoring inside is covered, we mark the ``Others' with the superscripts.
}
\end{tablenotes}
\end{table}

Table~\ref{tab:type} shows the statistics of the bugs found across different refactoring types. To have a clearer and more organized view of various refactoring, we merge and categorize these concrete refactoring types into higher-level categories. Specifically, the ``Refactoring'' column gives the concrete refactoring types with the corresponding total number of bug reports. All these refactoring types are supported by our studied refactoring engines. We further merge some common refactoring types into ``Subcategory'' based on their operations (e.g., extract, inline, move, etc.) and the granularity of the program to be refactored (e.g., class, method, and variable). Finally, refactorings with the same operations are grouped into the ``Category'' column, which gives the highest-level refactoring category along with the number and percentage of bug reports.

From a high-level perspective, as shown in the ``Category'' column in Table~\ref{tab:type}, ``Extract'' (149/28.76\%), ``Inline'' (88/16.99\%), and ``Move'' (80/15.44\%) are the top three most error-prone refactoring transformations, among which the number of bug reports for the ``Extract'' refactoring nearly double the number of ``Inline'' and ``Move'' refactoring. During manual analysis, we deduce two reasons for the error-prone refactoring types: (1) popularity of refactoring (our result is in line with prior studies~\cite{negara2013comparative,paixao2020behind,alomar2022refactoring,golubev2021one} that revealed that extract, inline, and move refactoring are among the most commonly applied operations). 
(2) those refactoring types are more general and serve various purposes (e.g., facilitate reusability, improve readability, remove duplication, etc.)~\cite{silva2016we,paixao2020behind,alomar2022refactoring}, thus they can be performed in some more complicated usage scenarios (e.g., extract a static method from a lambda expression inside an inner class~\cite{extractMethodRefactoringIDEAIssue354122}), which makes it more challenging for the refactoring engines to take into account all the complicated usage scenarios. For the concrete refactoring types supported by our studied refactoring engines in the ``Refactoring'' column, ``Inline Method'' (46), ``Move Method'' (42), and ``Extract Method'' (40) trigger the top three number of bugs, following by ``Change Method Signature'' (38), and ``Extract Variable'' (37). Developers are more familiar with these refactoring types and more likely to automate them with the support of IDEs according to a previous study~\cite{golubev2021one} concerning the refactoring usage conducted by the JetBrains IDE development teams. Although most bug reports in our study only involve one refactoring operation, five bug reports mention that the same bug appeared in more than one refactoring type (e.g., ``Extract variable'' and ``Extract Constant'' do not handle ``Text Block'' in \eclipse 4.12~\cite{multipleRefactoringEclipseIssue551002}) so we label them as ``multiple'' and categorize them into the ``Others'' row in the ``Refactoring'' column. For some refactoring types (e.g., ``Infer Type Arguments'' and ``Make Static'') which cannot be categorized into a more general category, we put them in the ``Others'' category. Among the refactoring types in ``Others'', if the number of bug reports is smaller than five (e.g., ``Surround With Try/Catch''), we merge them into the last row named ``Others'' in the ``Refactoring'' column.






\noindent\textbf{Refactoring types are covered by existing testing approaches.} We further analyze whether existing approaches ~\cite{daniel2007automated,soares2009saferefactor,mongiovi2016scaling,soares2010making,soares2012automated,soares2011identifying,soares2009generating,gligoric2013systematic} for testing refactoring engines can be used to automatically identify bugs for each refactoring type. Our analysis revealed that existing approaches only cover part of the refactoring types in Table~\ref{tab:type}. \astgen~\cite{daniel2007automated} focuses on testing only eight refactoring types (marked with a superscript `A' in Table~\ref{tab:type} ``Refactoring'' column) and the top three most error-prone refactoring types have not been included in the supported refactoring types. This is because \astgen relies on manually designed program generators for each refactoring type, which requires domain knowledge and incorporating various characteristics of input programs (we study and answer this question in RQ6) to effectively trigger the refactoring engine bugs. Several approaches based on \saferefactor~\cite{soares2009saferefactor,mongiovi2016scaling,soares2010making,soares2012automated,soares2011identifying,soares2009generating} also cover a limited number of refactoring types (marked with a superscript `S'). Meanwhile, Gligoric et al.~\cite{gligoric2013systematic} proposed an approach that supports testing of 23 refactoring types (marked with a superscript `G') available in the \eclipse 4.2 refactoring menu, however, they are still limited in (1) Some applicable elements for each refactoring are not covered. For example, they test instance method and static method for the move refactoring but there are other elements like class, constant, and field that are currently out of their scope; (2) Refactoring types that are unique in \idea and \netbeans (e.g., ``Convert To Enhanced For Loops'') are not covered because they only test \eclipse; (3) Refactoring types added in the newer version (after version 4.2) of \eclipse are not covered (e.g., ``Convert String to Text Block''). In total, ~\coveredrefactoringtype{} of the ~\totalrefactoringtype{} refactoring types in Table~\ref{tab:type} are covered by at least one of the existing refactoring testing tools. As the remaining refactoring types have not been covered, developers of refactoring engines for these unsupported types have to resort to testing them manually. This indicates a gap in existing testing tools and the need to design a general testing approach that can support more refactoring types.



\begin{tcolorbox}[left=0pt,right=0pt,top=0pt,bottom=0pt]
\textbf{Finding 1:} ``Extract'' (149/28.76\%), ``Inline'' (88/16.99\%), and ``Move'' (80/15.44\%) are the top three most error-prone refactoring types. Existing testing tools only cover about one-third (\coveredrefactoringtype{}/\totalrefactoringtype{}) of the refactoring types.   
\end{tcolorbox}

\subsection{RQ2: Bug Symptoms}
\label{sec:motivation}

Based on the classification and labeling process described in Section~\ref{sec:classification}, we identify several symptoms of refactoring engine bugs below:





\noindent\textbf{Compile Error (242).} The refactored program is syntactically incorrect, resulting in compile errors.

\noindent\textbf{Crash (106).} The refactoring engine terminates unexpectedly during refactoring, which usually produces an error message.

\noindent\textbf{\behaviorchange{} (66).} The refactoring engine successfully applies the corresponding refactoring and produces a refactored program without any syntax error. However, the refactored program is not behavior-preserving, which violates the definition of refactoring~\cite{eclipseBehaviorChangeExample}.




\noindent\textbf{Incorrect Warning Message (24).} The refactoring engine gives an incorrect or confusing warning message to prevent the refactoring from being performed. For example, applying ``Change Method Signature'' refactoring in \eclipse 4.17 for a method in a \textit{Record}~\cite{javarecordclass} type declaration would raise a warning message ``could not resolve type int''. This warning message is incorrect as the developer states ``continue renames correctly, but the error is unwarranted''~\cite{eclipseIncorrectWarningMessageExample}.

\noindent\textbf{Failed Refactoring (23).} The refactoring engine either makes no change to the original program or only partially completes the refactoring. In this case, the refactored program is still compilable and behavior-preserving, but the result does not align with the user's intention since the refactoring either changes nothing at all or has only been partially completed. 
For example in \eclipse 3.1, if there are two identical expressions in a program~\cite{eclipseFailToRefactoringExample} when a user performs ``Extract Local Variable'' on one expression and selects the ``Replace All Occurrences'' checkbox, \eclipse only replaces the selected expression with the extracted variable but the other identical expression remains unchanged. The refactored program could be compiled successfully but the refactoring result does not align with the user's intention as the users would like all occurrences of the expression to be refactored correctly.

\noindent\textbf{Comment Related (22).} The refactoring engine successfully applies the corresponding refactoring type and produces a refactored program without any syntax error. However, the refactoring engine fails to resolve code comments appropriately (e.g., deleted comments~\cite{eclipseDeleteCommentExample}, creating wrong comments~\cite{ideaCreatewrongCommentExample}, and broken Javadoc comments~\cite{ideaBrokenCommentExample}).

\noindent\textbf{Unnecessary Change (17).} The refactoring engine successfully applies the corresponding refactoring and produces a refactored program without any syntax error. However, the refactored programs contain some redundant changes (e.g., unnecessary parentheses~\cite{ideaRedundantChangeUnnecessaryParenthesesExample}). This happens because the refactoring engine developers are over-cautious, thus they make verbose changes (adds code that is not needed) to ensure the refactored program will not introduce errors. Even though those redundant changes will not introduce a compilation error or change the program behavior, the refactoring result is not aligned with the developer's expectations.

\noindent\textbf{Refactoring Not Available (12).} The refactoring operation is not available on certain program elements (e.g., text block and annotation) even though it is supposed to be. For example, in \eclipse 4.12, the user complained that ``Extract Local Variable'', ``Extract Constant'', and ``Inline'' refactorings are not available on ``Text Block''~\cite{notavailableexample}.

\noindent\textbf{Others (6).} The remaining bugs show other specific symptoms, e.g., broken breakpoints, and bad performance.





\begin{table}
\centering
\caption{Bug Distribution by Symptoms}
\label{tab:symptom}
\begin{adjustbox}{width=0.6\textwidth}
\begin{tabular}{l|rrr|r} 
\toprule
\textbf{Symptom}          & \textbf{Eclipse} & \textbf{IntelliJ IDEA} & \textbf{NetBeans} & \textbf{Total}  \\ 
\hline
Compile Error             & 88               & 114                    & 40                & 242             \\
Crash                     & 82               & 14                     & 10                & 106             \\
Behavior Change           & 23               & 33                     & 10                & 66              \\
Failed Refactoring        & 9                & 10                     & 5                 & 24              \\
Incorrect Warning Message & 9               & 14                     & 0                 & 23              \\
Comment Related           & 7                & 12                     & 3                 & 22              \\
Unnecessary Change        & 7                & 9                     & 1                 & 17              \\
Refactoring Not Available & 9                & 3                      & 0                 & 12              \\
Others                    & 6                & 0                      & 0                 & 6               \\
\bottomrule
\end{tabular}
\end{adjustbox}
\end{table}

Table~\ref{tab:symptom} presents the distribution of refactoring engine bugs according to the symptom categories. It shows that ``Compile Error'' is the most common symptom in all three refactoring engines. It occurs when refactoring engines generate refactored programs containing syntax errors, accounting for 88 \eclipse bugs, 114 \idea bugs, and 40 \netbeans bugs, respectively. Determining the syntactic correctness of a refactored program is not difficult with the help of JVM compilers. Therefore, the developer could test such bugs by compiling the refactored program and collecting the error messages to fix the syntax errors. ``Crash''  is the second most common symptom. 82, 14, and 10 bugs exhibit this symptom in \eclipse, \idea, and \netbeans, respectively. As detection of crashes does not require explicit test oracles, the large percentage of crashes suggests the potential of augmenting the existing test suite with the generated or mutated ones. Besides, engine developers can design effective localization and deduplication methods based on the stack trace information collected from crashes to locate the root causes of the crashes. ``Behavior Change'' occurs when refactoring engines generate non-behavior-preserving refactored code. It is the third common symptom as shown in Table~\ref{tab:symptom}, taking up 66 bugs in total. This symptom is not as obvious as ``Compile Error'' and ``Crash''. Specifically, the refactored program contains no syntax errors, determining the behavior-preserving of a refactoring is hard due to its complexity. Existing testing tools based on \saferefactor~\cite{soares2009saferefactor,mongiovi2016scaling,soares2010making,soares2012automated,soares2011identifying,soares2009generating} try to identify those kind of bugs by randomly generating tests, however, they can only generate test cases for the common methods before and after the refactoring in the project, which means they cannot exercise the changed methods (e.g., refactored methods) thoroughly. Besides, according to a previous study~\cite{silva2017analyzing}, automatic test generation tools (e.g., Randoop
and EvoSuite) are not effective for refactoring validation and they could miss more than half of injected refactoring faults. Therefore, testing such bugs is challenging because the test oracles are difficult to define. Further, the adverse impact of this category of bugs is severe since it violates the behavior-preserving assumption of a refactoring engine. Hence, our study indicates that it is worthwhile for future research in testing refactoring engines to focus on designing effective techniques for identifying non-behavior-preserving refactoring bugs.


\begin{tcolorbox}[left=0pt,right=0pt,top=0pt,bottom=0pt]
\textbf{Finding 2:} ``Compile Error'' (242),  ``Crash'' (106), and ``\behaviorchange'' (66) are the top three most common symptoms of refactoring engine bugs. Designing test oracles for identifying ``\behaviorchange'' bugs is challenging.
\end{tcolorbox}

According to Table~\ref{tab:symptom}, the symptoms of ``Failed Refactoring'' (24), ``Incorrect Warning Message'' (23), and ``Comment Related'' (22) are non-negligible. ``Failed Refactoring'' could be partly detected by checking whether the original program has been changed or not, but for the partially completed refactoring, it is hard to detect the incomplete change since it still compiles successfully without any syntax error. Detection of ``Incorrect Warning Message'' and ``Comment Related'' is also difficult since there are no standard oracles, but with multiple refactoring engines, differential testing which has been used in testing static analysis tools~\cite{zhang2024understanding} could be used to compare the refactoring results to find the defects. Similarly, testing ``Unnecessary Change'' (17) is challenging because the test oracles are difficult to define (the question about ``which change is redundant?'' needs to be answered to design the oracle for each refactoring type) and one possible solution is to use the refactoring results of multiple refactoring engines to identify the redundant change. Meanwhile, ``Refactoring Not Available'' (12) could be detected by checking for the availability of the refactoring type for the input programs that satisfy the preconditions. Meanwhile, we classify some symptoms into ``Others'' since their bug number is small, like the ``Bad performance'' (2) and ``Broken Breakpoints'' (3). The ``Broken Breakpoints'' together with the ``Comment Related'' symptoms indicate that developers of refactoring engines should take into account the various side effects of the refactoring operations because maintaining the consistency between the refactored code and other software artifacts (e.g., code comments and breakpoints) is as important as preserving the functionality of the original program.

\begin{tcolorbox}[left=0pt,right=0pt,top=0pt,bottom=0pt]
\textbf{Finding 3:} Apart from the refactored code itself, other aspects such as warning messages and refactoring availability could also be error-prone. Current refactoring engines have not thoroughly taken into account various side effects that may be incurred by the refactoring. It is important to maintain the consistency between the refactored code and other software artifacts (e.g., code comments, and breakpoints).

\end{tcolorbox}

\subsection{RQ3: Root Causes}
\label{sec:auto}

Based on the above classification and labeling process, all the root causes of refactoring engine bugs are presented as follows.

\noindent\textbf{Incorrect Transformations (165).} The most common root cause is incorrect transformations, this happens when refactoring engines try to rewrite the AST and create code modifications. Due to the transformations in the refactoring engines being rule-based and manually crafted which heavily rely on the refactoring engine developers' expertise, it is natural to be error-prone. Besides, the developers are not aware of all the scenarios for each refactoring considering the diversity and grammar complexity of the input programs. For example, in Figure~\ref{lst:eclipseIncorrectTransformation}, when a user performs ``Inline Method'' refactoring for ``originalMethod()'' in line 10 using \eclipse version 2023-09, the keyword ``synchronized'' is missed in the refactored code leading to a behavior change, as the user states: ``Before refactoring, inline methods contain access permissions for synchronization modifiers, but are missing after inlining.''~\cite{eclipseIncorrectTransformationExample} By analyzing the patches together with the commit messages, we conclude this happens because refactoring engine developers ignore this kind of situation in the input program, and thus do not add logic to process the synchronized block~\cite{eclipseIncorrectTransformationPatchExample}.

\begin{figure}
\caption{Inline method refactoring leads to changes in access permissions}
\label{lst:eclipseIncorrectTransformation}
{\centering
\begin{adjustbox}{width=0.4\textwidth}

\begin{lstlisting}[style=mystyle, escapechar=^]
class OriginalClass {
     private boolean flag = false;
-    public synchronized void originalMethod(){
+    public void callerMethod(){
         // Some logic here
         flag = true;
         notify();
     }
-    public void callerMethod(){
-        originalMethod();
-    }
}
\end{lstlisting}\par
\end{adjustbox}\par
} 
\end{figure}

\noindent\textbf{Incorrect Preconditions Checking (62).} Each refactoring contains some preconditions to guarantee the refactored program to be behavior-preserving or syntactic correct. For example, when implementing \makestatic{} refactoring, the \eclipse developers propose a list of preconditions that need to be checked~\cite{makeStaticRefactoring} (e.g., the method to be made static should not be a constructor method). In practice, testing refactoring preconditions involves manually creating an input program to be refactored and specifying a refactoring precondition failure as expected output. However, developers choose input programs for checking just the preconditions they are aware of. Since specifying preconditions is a non-trivial task, developers may be unaware of the preconditions needed to guarantee behavioral preservation or syntax correction. When the implemented preconditions are insufficient to guarantee behavioral preservation or syntactic correctness, we call it ``Overly Weak Preconditions'', which may introduce compilation errors or behavioral changes. Additionally, some implemented preconditions may be overly strict, we define them as ``Overly Strong Preconditions'', thus it cause the engines to prevent developers from applying applicable transformations. ``Overly Weak Preconditions'' and ``Overly Strong Preconditions'' account for 55 and 9 bugs, respectively.

\noindent\textbf{Incorrect Flow Analysis (51).} Incorrect flow analysis in refactoring engines is a common bug root cause in refactoring that depends on the code's control flow and data flow. Refactoring like the ``Extract Method'' often relies heavily on accurate flow analysis to identify variables, control structures, and dependencies correctly, when the flow analysis is flawed or not in the right scope, it can lead to erroneous refactoring results. For example, in Figure \ref{lst:ideaIncorrectFlowAnalysis}~\cite{ideaIncorrectFlowAnalysisExample}, \idea fails to detect input variables when the code fragment to be extracted is located in an unreachable else branch, resulting in an uncompilable refactored code. This issue happens because \idea skips the unreachable code during control flow analysis for the ``Extract Method'' refactoring~\cite{ideaIncorrectFlowAnalysisExamplePatch}.


\begin{figure}
\caption{\idea fails to detect input variable if code fragment is located in unreachable else branch, the extracted method contains syntax error}
\label{lst:ideaIncorrectFlowAnalysis}
{\centering
\begin{adjustbox}{width=0.4\textwidth}

\begin{lstlisting}[style=mystyle, escapechar=^]
public class Test {
    public void test() {
        int a = 1;
        if (true) {
            System.out.println(a);
        } else {
-            System.out.println(a);
+            foo();
        }
    }
+    private void foo() { 
+        System.out.println(a);//syntax error
+    }
}
\end{lstlisting}\par
\end{adjustbox}\par
} 
\end{figure}

\noindent\textbf{Incorrect Type Resolving (35).} Bugs in this category is caused by the incorrect type-related operations such as type inference, type binding, type matching, and etc. For instance, type binding refers to the process of resolving references in Java source code to their corresponding types. Specifically, a type binding associates variable declarations, method parameters, or return types with their concrete types. This information is stored in the AST's binding nodes for further operations. The bindings are crucial for understanding the program's structure and ensuring that the IDEs can perform accurate code analysis, such as renaming symbols.


\noindent\textbf{Failed Selection Parsing (9).} This category of bugs occurs when the refactoring engine fails to parse the input programs. According to the discussions in the bug reports and the patches, we further divide this category into three subcategories:
\begin{itemize}

\item{\textbf{Unsupported new Java language features (4).}} This happens because some language features in the newest version of JDK are not fully supported in the IDEs. For instance, in \eclipse bug report \#156~\cite{eclipseFailedParsingExample}, a user reports that rename refactoring does not work for ``Record Patterns''~\cite{recordPatterns}, as the refactoring engine developers reply in the bug report: ``Currently, there is no AST node for variables like xy and lr. This should be revisited after Record Pattern DOM AST changes is merged''.

\item{\textbf{Parsing for some program elements is not well-designed (3).}} For example, \idea users complain that some refactorings do not work for strings containing injected languages~\cite{ideaInjectedLanguageExample}, or file paths~\cite{ideaStringPathExample}. The same situation also exists in \eclipse, as shown in Figure~\ref{lst:eclipsefailtoextractconstant}, when users try to extract a constant for a string that is the value for annotation, the refactoring is not available on the pop-up because annotation type is not implemented when parsing Java text selection~\cite{eclipseExtractConstantFromAnnotationExample} to extract.

\begin{figure}
\caption{Refactoring is not available on pop-up if selected string is in an annotation in \eclipse 3.6}
\label{lst:eclipsefailtoextractconstant}
{\centering
\begin{adjustbox}{width=0.4\textwidth}

\begin{lstlisting}[style=mystyle, escapechar=^]
import java.util.ArrayList;
//@SuppressWarnings("all")
//@SuppressWarnings({"all", "rawtypes"})
//@SuppressWarnings(value= "all")
@SuppressWarnings(value= {"all", "rawtypes"})
public class Try extends ArrayList {
}
\end{lstlisting}\par
\end{adjustbox}\par
} 
\end{figure}

\item{\textbf{Inflexible parsing (2).}} Some users complain that refactoring engines are not flexible when parsing the selected program elements to be refactored. For example, ``Extract Method'' refactoring does not work when a comment is selected at the end of the statements in \idea~\cite{ideaInflexibleParsingExample}. In \eclipse 3.8, ``Extract Method'' refactoring would fail if the trailing comma is not selected for a statement~\cite{eclipseInflexibleParsingExample}.
\end{itemize}

\noindent\textbf{Others (4).} Each case in this root cause occurs not frequently and does not belong to any other root causes above. Including compatibility issues with the JDK version~\cite{eclipseJdkExample}, performance issues caused by time-consuming method calls (e.g., the ``toString()'' method might cause a loss of performance~\cite{eclipseToString}), and incorrectly getting user configuration~\cite{ideaInitializationIssueExample}.

\begin{figure}[h]
    \centering
    \vspace{-3pt}    \includegraphics[width=0.6\textwidth]{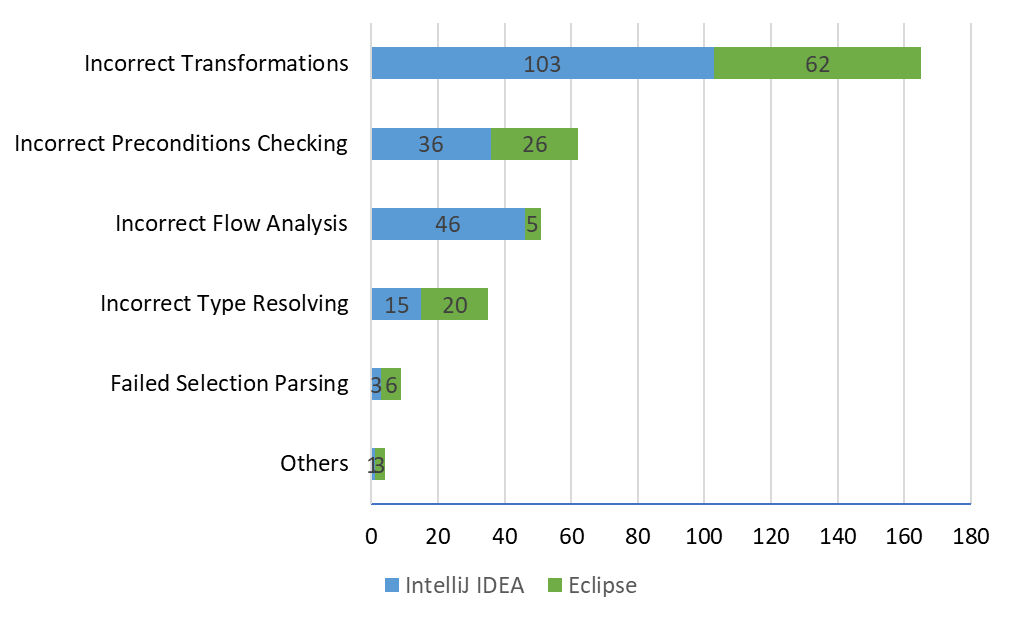}
    \vspace{-6pt}
     \caption{Bug Distribution by Root Causes}
    \label{bugDistributionByRootCause}
\end{figure}

Figure~\ref{bugDistributionByRootCause} gives the distribution of bugs by their root causes. As shown in the figure, ``Incorrect Transformations'' is the most common root cause. It accounts for 165 bugs in total, including 103 bugs in \idea, and 62 bugs in \eclipse. This happens because refactoring engine developers need to write the code transformation rules manually depending on the input programs to be refactored. However, the input programs could be diverse due to the complex grammar and different usage scenarios, thus developers might not be aware of the input program beyond their knowledge, as a result, the designed rules could be flawed. On the other hand, developers could make mistakes even with the input programs since writing transformation rules manually is non-trivial. The result indicates that handling transformation issues in refactoring engines is very challenging and deserves more attention. Based on our study, we conclude two directions that developers and researchers could focus on to tackle this problem:

\begin{itemize}

\item{\textbf{Towards obtaining diverse input programs.}} Diverse input programs could help developers from two aspects: First, those input programs could be used to design unit tests to expose bugs in refactoring engines; Second, with the help of diverse input programs, developers could design more robust program transformation rules considering different refactoring scenarios. The diverse input programs could be obtained from real-world projects with the help of automated refactoring detection tools (e.g., RefactoringMiner~\cite{refactoringminerurl}). Alternatively, since large language models (LLMs) like ChatGPT trained on massive data have achieved impressive results in many software engineering tasks, leveraging LLMs to generate diverse input programs could be promising. Historical refactoring engine bug reports or input programs in the existing test suites could be used to construct the prompts to generate more diverse input programs.

\item{\textbf{Automate the code refactoring in a data-driven way.}} Previous studies~\cite{li2024multilingual,liu2018deep,liu2023deep,liu2019deep} have tried to automate the code transformations in a data-driven way. They have obtained comparable results, besides, more diverse transformations could be applied since they are trained on massive data from real-world projects. Recently, \idea has introduced code AI assistants to help users automate code-related tasks, including suggest refactoring~\cite{ideaAIAssistant}. It could be potential to explore code refactoring automation in the era of LLMs.

\end{itemize}

\begin{tcolorbox}[left=0pt,right=0pt,top=0pt,bottom=0pt]
\textbf{Finding 4:} Our study found that ``Incorrect Transformations'' (165) is the most common root cause for both \eclipse and \idea, accounting for 62 and 103 bugs, respectively. We conclude two directions as resolution: (1) Towards obtaining more diverse input programs; (2) Automate the code refactoring in a data-driven way.
\end{tcolorbox}

``Incorrect Preconditions Checking'' is the second most common root cause, accounting for 62 bugs, including 54 bugs caused by ``Overly Weak Preconditions'' and 8 bugs caused by ``Overly Strong Preconditions''. Existing studies~\cite{soares2011identifying,mongiovi2017detecting} mainly focus on revealing the bugs caused by ``Incorrect Preconditions Checking'', however, as shown in Figure~\ref{bugDistributionByRootCause}, it only takes up a small portion of bugs. Besides, manually crafting preconditions for each refactoring could be non-trivial since developers need to consider different scenarios of input programs. Meanwhile, with the introduction of new language features in the newer version of JDK, each refactoring preconditions need to be updated accordingly. There are works~\cite{astorga2018preinfer,padhi2016data,cousot2013automatic,kafle2021transformation} that try to infer preconditions for debugging or program verification through symbolic analysis or in a data-driven way. However, the precondition inference for refactoring has not been well-explored. Given the input program and existing preconditions, new preconditions should be predicted automatically.

\begin{tcolorbox}[left=0pt,right=0pt,top=0pt,bottom=0pt]
\textbf{Finding 5:} ``Incorrect Preconditions Checking'' (62) is the second most common root cause. Preconditions need to be updated considering the introduction of new language features. Considering the evolution of programming language, automating the inference of preconditions for refactoring could be a promising research direction.
\end{tcolorbox}

The bugs caused by ``Incorrect Flow Analysis'', and ``Incorrect Type Resolving'' are also non-negligible, accounting for 51 bugs and 35 bugs, respectively. Flow analysis is fundamental for refactoring, ineffective flow analysis could result in behavior change, failed refactoring, and uncompilable code. A previous study~\cite{chi2023automated} also has shown that ``Extract Local Variable'' refactoring in \eclipse and \idea could lead to behavior change because of inaccurate flow analysis and ineffective side effects identification. ``Incorrect Type Resolving'' (i.e., incorrect handling of type-related operations), such as type inference, type binding, and type matching, is a common root cause of bugs. These issues often arise because refactoring involves complex code transformations that depend on correctly manipulating the types of variables, expressions, and methods. For instance, during refactoring, especially in Java versions that support features like the ``var'' keyword, refactoring may fail because of incorrectly inferred types. For example, \eclipse 4.11 fails to convert a local variable to a field on the ``var'' variable because it does not get resolved to its inferred type during refactoring~\cite{eclipseVarExample}. Meanwhile, ``Failed Selection Parsing'' (16) caused by some rare input programs, unsupported language features, and inflexible parsing should also be paid attention. Except for the above-mentioned root causes, refactoring engines might be affected by the different versions of JDK, or slowed down by some time-consuming method calls.

\begin{tcolorbox}[left=0pt,right=0pt,top=0pt,bottom=0pt]
\textbf{Finding 6:} ``Incorrect Flow Analysis'' (51), and ``Incorrect Type Resolving'' (35) are also non-negligible root causes. Refactoring engine developers should take into account different scenarios for parsing input programs.
\end{tcolorbox}

\subsection{RQ4: Root Causes and Symptoms}

\begin{table}
\centering
\caption{Relationship between Root Causes and Symptoms}
\label{tab:relationship}
\begin{adjustbox}{width=0.98\textwidth}
\begin{tabular}{c|rrrrrrrrr|r} 
\toprule
\diagbox{\textbf{Root Causes}}{\textbf{Symptoms}} & \textbf{CE}  & \textbf{Crash} & \textbf{BC} & \textbf{IWM} & \textbf{FR} & \textbf{CR} & \textbf{UC} & \textbf{RNA} & \textbf{Others} & \textbf{Total}  \\ 
\hline
Incorrect Transformations                         & 80  & 27    & 21 & 4   & 4  & 15 & 11 & -   & 3      & 165             \\
Incorrect Preconditions Checking                  & 28  & 15    & 9  & 7   & 2  & -  & -  & -   & 1      & 62              \\
Incorrect Flow Analysis                           & 27  & 7     & 6  & 7   & 3  & -  & 1  & -   & -      & 51              \\
Incorrect Type Resolving                          & 14  & 8     & 7  & -   & 6  & -  & -  & -   & -      & 35              \\
Failed Selection Parsing                          & -   & 3     & -  & -   & -  & -  & -  & 6   & -      & 9               \\
Others                                            & 1   & 1     & -  & -   & -  & -  & -  & 1   & 1      & 4               \\ 
\hline
\textbf{Total}                                    & 150 & 61    & 43 & 18  & 15 & 15 & 12 & 7   & 5      & 326             \\
\bottomrule
\end{tabular}
\end{adjustbox}
\begin{tablenotes}
\footnotesize
\item{
\textbf{CE}: Compile Error; \textbf{BC}: Behavior Change; \textbf{IWM}: Incorrect Warning Message; \textbf{FR}: Failed Refactoring; \textbf{CR}: Comment Related; \textbf{UC}: Unnecessary Change; \textbf{RNA}: Refactoring Not Available.
}
\end{tablenotes}
\end{table}

Table~\ref{tab:relationship} presents the number of refactoring engine bugs caused by each root cause with each symptom. As the most common root cause, ``Incorrect Transformations'' occurs in almost all categories of symptoms (except ``Refactoring Not Available''). Other root causes that can result in a broad category of symptoms are ``Incorrect Preconditions Checking'' and ``Incorrect Flow Analysis''. The bugs caused by the above three root causes not only occur frequently but also produce a wide variety of effects. Moreover, the bugs introduced by ``Incorrect Transformations'' and ``Incorrect Preconditions Checking'' tend to be specific to refactoring engines. Therefore, more attention from both refactoring engine developers and researchers should be paid to detect, localize, and fix these bugs. The symptom of ``Incorrect Type Resolving'' is also non-negligible. ``Compile Error'', ``Crash'', and ``Behavior Change'' bugs are mainly caused by ``Incorrect Transformations'', and they are the most common symptoms. They can be induced by most of the root causes except two (i.e., ``Failed Selection Parsing'' and ``Others''). Therefore, generating high-quality test oracles around the three symptoms can detect a wide variety of bugs.

\begin{tcolorbox}[left=0pt,right=0pt,top=0pt,bottom=0pt]
\textbf{Finding 7:} ``Incorrect Transformations'' can induce all kinds of buggy symptoms except for ``Refactoring Not Available'', which is mostly exhibited by ``Failed Selection Parsing''. ``Incorrect Transformations'', ``Incorrect Preconditions Checking'', and ``Incorrect Flow Analysis'' are the common root causes for the top three symptoms.
\end{tcolorbox}

\subsection{RQ5: Triggering Conditions}
\label{sec:triggering}

We check the discussions and procedures to reproduce bugs in each bug report. If a user does not explicitly mention changing the default refactoring input options, we consider the bug could be triggered by the default configuration. Overall, only 15 (2.9\%) bugs are not triggered by the default configuration, in those cases, users need to select extra options in the refactoring user interfaces (e.g., when extract method in \idea, user could check the ``Declare static'' to make the extracted method as a static method). This means the rest of the refactoring engine bugs (97.1\%) could be triggered by the default initial input options. This is a useful finding since configuring the input options while testing refactoring engines could increase search space and is time-consuming. For example, while using fuzzing testing to generate input programs to uncover the bugs, we could just use the default input options without considering the configuration-related bugs, which could be more faster and efficient.

\begin{tcolorbox}[left=0pt,right=0pt,top=0pt,bottom=0pt]
\textbf{Finding 8:} Most of the bugs (97.1\%) could be triggered by the default initial input options of refactoring engines. 
\end{tcolorbox}

For the \refactorbench dataset, we manually check the input program and categorize its characteristics. This step aims to identify input program characteristics that are more prone to trigger refactoring engine bugs. \emph{The purposes of the current research question are: 1) By identifying the input program characteristics and features, the researchers and engine developers could use that information to guide the test case generation techniques to produce more error-prone input programs, improving the effectiveness and efficiency of testing. For example, researchers could use our findings in Table~\ref{tab:characteristics} and seed input programs from our dataset to construct prompts and instruct the LLMs (e.g., ChatGPT) to generate more diverse input programs. In this way, it could potentially trigger more bugs in the refactoring engines. 2) The engine developers could design more diverse and complete unit tests for each refactoring type by referring to our findings, improving the IDEs' robustness and usability.} We use the methodology mentioned in Section 3.2 to derive the taxonomy.

\begin{table*}
\def\arraystretch{1.0}
\centering
\caption{Categories of Input Program in Our Studied Bug Reports}
\label{tab:characteristics}
\begin{adjustbox}{width=0.98\textwidth}
\begin{tabular}{lllcc} 
\toprule
\multicolumn{1}{c}{Category}              & \multicolumn{1}{c}{Sub-category}                           & Description                                                                                                                                                               & Bug (\#)             & Total (\#/\%)                        \\ 
\hline
\multirow{17}{*}{(T1) Language Features}  & (T1.1) Lambda expression                                   & \textcolor[rgb]{0.051,0.051,0.051}{Anonymous functions used to implement functional interfaces with a more streamlined syntax}                                            & \textbf{39}          & \multirow{17}{*}{\textbf{174/64.4}}  \\
                                          & (T1.2) Java generics                                       & \textcolor[rgb]{0.051,0.051,0.051}{Java generics allow to create classes, interfaces, and methods that operate with unspecified types}                                    & \textbf{\textbf{38}} &                                      \\
                                          & (T1.3) Enum                                                & \textcolor[rgb]{0.051,0.051,0.051}{A special data type used to define a fixed set of constants representing a set of predefined values}                                   & \textbf{\textbf{27}} &                                      \\
                                          & (T1.4) Record                                              & A special kind of class that helps encapsulating related data, introduced in~Java 14                                                                                      & 13                   &                                      \\
                                          & (T1.5) Varargs                                             & \textcolor[rgb]{0.051,0.051,0.051}{It allows a method to accept a variable number of arguments of the same type}                                                          & 11                   &                                      \\
                                          & (T1.6) instanceof                                          & \textcolor[rgb]{0.051,0.051,0.051}{It}\textcolor[rgb]{0.051,0.051,0.051}{~is a Java keyword used to check if an object is an instance of a particular class or interface} & 8                    &                                      \\
                                          & (T1.7) Foreach                                             & \textcolor[rgb]{0.051,0.051,0.051}{A concise loop construct used to iterate over elements in an array or collection without explicit indexing}                            & 8                    &                                      \\
                                          & (T1.8) Switch case                                         & \textcolor[rgb]{0.051,0.051,0.051}{It provides a concise way to conditionally execute blocks of code based on the value of a variable}                                    & 8                    &                                      \\
                                          & (T1.9) Try-with-resources                                  & \textcolor[rgb]{0.051,0.051,0.051}{It can that automatically manages resources, ensuring they are closed at the end of a block of code}                                   & 5                    &                                      \\
                                          & (T1.10) Var                                                & \textcolor[rgb]{0.051,0.051,0.051}{Enables implicit type inference, allowing for more concise code while maintaining strong typing}                                       & 4                    &                                      \\
                                          & (T1.11) Try-catch-finally                                  & \textcolor[rgb]{0.051,0.051,0.051}{Java try-catch-finally is a mechanism used for handling exceptions}                                                                    & 3                    &                                      \\
                                          & (T1.12) Joint variable/field declaration                   & \textcolor[rgb]{0.051,0.051,0.051}{It allows multiple variables or fields of the same type to be declared in a single statement}                                          & 3                    &                                      \\
                                          & (T1.13) Multi-dimension Array                              & \textcolor[rgb]{0.051,0.051,0.051}{A data structure that organizes elements in multiple dimensions}                                                                       & 2                    &                                      \\
                                          & (T1.14) Vector                                             & \textcolor[rgb]{0.051,0.051,0.051}{Vector is a synchronized, resizable array implementation of the List interface}                                                        & 2                    &                                      \\
                                          & (T1.15) Synchoronized block                                & \textcolor[rgb]{0.051,0.051,0.051}{It ensures that only one thread at a time can execute the code within the block}                                                       & 1                    &                                      \\
                                          & (T1.16) Java ternary conditional                           & \textcolor[rgb]{0.051,0.051,0.051}{A concise way to write conditional statements, returning one of two values based on a specified condition}                             & 1                    &                                      \\
                                          & (T1.17)~Keyword ``this''                                     & \textcolor[rgb]{0.051,0.051,0.051}{The ``this'' keyword in Java refers to the current instance of the class, allowing access to its own members}                            & 1                    &                                      \\ 
\hline
\multirow{3}{*}{(T2) Class-related}       & (T2.1) Inner class                                         & \textcolor[rgb]{0.051,0.051,0.051}{Class declared within another class or method, enabling tighter encapsulation and logical grouping of code}                            & 19                   & \multirow{3}{*}{\textbf{35/13.0}}    \\
                                          & (T2.2)~Anonymous class                                     & \textcolor[rgb]{0.051,0.051,0.051}{Class defined without a name, often used for one-time implementations of interfaces or abstract classes}                               & 15                   &                                      \\
                                          & (T2.3) Cyclically dependent class                          & \textcolor[rgb]{0.051,0.051,0.051}{Two or more classes depend on each other directly or indirectly}                                                                       & 1                    &                                      \\ 
\hline
(T3) Annotations                          & (T3.1)~Annotations                                         & \textcolor[rgb]{0.051,0.051,0.051}{Annotations can be used by the compiler or at runtime for various purposes such as configuration}                                      & 29                   & \textbf{\textbf{29/10.7}}            \\ 
\hline
(T4) Code Comment                         & (T4.1) Comment related                                     & Related with Java comments, including comments in method and class                                                                                                        & 18                   & 18/6.7                               \\ 
\hline
\multirow{5}{*}{(T5) Method-related}      & (T5.1) Overloaded method                                   & \textcolor[rgb]{0.051,0.051,0.051}{Methods within the same class that share the same name but have different parameter lists}                                             & 6                    & \multirow{5}{*}{12/4.4}              \\
                                          & (T5.2) Static method                                       & \textcolor[rgb]{0.051,0.051,0.051}{A method that belongs to the class rather than any specific instance}                                                                  & 2                    &                                      \\
                                          & (T5.3) Method reference                                    & \textcolor[rgb]{0.051,0.051,0.051}{A shorthand syntax for referring to methods by their names, enabling concise and readable code}                                        & 2                    &                                      \\
                                          & (T5.4) Recursive method                                    & \textcolor[rgb]{0.051,0.051,0.051}{A function that calls itself within its definition}                                                                                    & 1                    &                                      \\
                                          & (T5.5)~Default method                                      & \textcolor[rgb]{0.051,0.051,0.051}{Method defined within an interface with a default implementation}                                                                      & 1                    &                                      \\ 
\hline
\multirow{3}{*}{(T6) Static}              & (T6.1) Static initializer                                  & \textcolor[rgb]{0.051,0.051,0.051}{It is typically used for initializing static fields or performing one-time setup tasks}                                                & 4                    & \multirow{3}{*}{7/2.6}               \\
                                          & (T6.2) Static import                                       & \textcolor[rgb]{0.051,0.051,0.051}{Fields and methods defined in one class to be referenced in another class without specifying the class name}                           & 2                    &                                      \\
                                          & (T6.3) Static field                                        & \textcolor[rgb]{0.051,0.051,0.051}{A class-level variable shared among all instances of the class}                                                                        & 1                    &                                      \\ 
\hline
\multirow{3}{*}{(T7) Constructor-related} & (T7.1) Super constructor                                   & \textcolor[rgb]{0.051,0.051,0.051}{It is used to initialize the superclass of an object}                                                                                  & 4                    & \multirow{3}{*}{6/2.2}               \\
                                          & (T7.2) Nested constructor                                  & \textcolor[rgb]{0.051,0.051,0.051}{Refers to a constructor within a class that can be invoked by another constructor in the same class}                                   & 1                    &                                      \\
                                          & (T7.3)~Implicit constructor                                & \textcolor[rgb]{0.051,0.051,0.051}{An implicit constructor is automatically provided by the compiler when no explicit constructor is defined}                             & 1                    &                                      \\ 
\hline
\multirow{5}{*}{(T8) Others}              & (T8.1) Special String                                      & Like injected languages, string contain special tokens                                                                                                                    & 6                    & \multirow{5}{*}{11/4.1}              \\
                                          & (T8.2) Arithmetic expression                               & \textcolor[rgb]{0.051,0.051,0.051}{Mathematical operations performed on numerical values using operators like +, -, *, /, and \%}                                         & 2                    &                                      \\
                                          & (T8.3) Time-consuming method call                          & For example, calling toString() method in a instance involving~\textcolor[rgb]{0.051,0.051,0.051}{resource access~}would be time-consuming                                & 1                    &                                      \\
                                          & (T8.4) Dead code block                                     & \textcolor[rgb]{0.051,0.051,0.051}{A block of code that is never executed}                                                                                                & 1                    &                                      \\
                                          & (T8.5)~\textcolor[rgb]{0.051,0.051,0.051}{Method chaining} & \textcolor[rgb]{0.051,0.051,0.051}{A sequence of operations on an object without needing to store intermediate results in separate variables}                             & 1                    &                                      \\
\bottomrule
\end{tabular}
\end{adjustbox}
\end{table*}

Table~\ref{tab:characteristics} lists the identified taxonomy of input program characteristics. Note that input programs in some bug reports have no explicit features, and there could be multiple characteristics existing in one bug report. Finally, we derive our taxonomy based on 270 bug reports (124 from \eclipse, 121 from \idea, and 25 from \netbeans). The ``Category'' column in Table~\ref{tab:characteristics} describes the high-level types of program characteristics, while the ``Sub-category'' column gives the specific categories. The last column (Total (\#/\%)) presents the total number and percentage of reports that fit into a particular category. In total, we identified \inputProgramCategoryNumber{} main categories with \inputProgramSubcategoryNumber{} sub-categories. Among the \inputProgramCategoryNumber{} categories, we observed that input programs involving Java language features are the most likely to trigger refactoring engine bugs during refactoring activities. For example, refactoring with lambda expression in the input program could trigger 39 bugs, since code transformation for lambda expression usually involves type inference and complex AST rewrite. This category takes up more than half (\inputProgramJavaLanguageFeatureRatio{}) of our studied bug reports. Existing studies ~\cite{daniel2007automated,soares2009saferefactor,mongiovi2016scaling,soares2010making,soares2012automated,soares2011identifying,soares2009generating,gligoric2013systematic} to test the refactoring engines are unaware of the bug-triggering ability of the input programs. ASTGen~\cite{daniel2007automated} relies on manually defined templates to generate input programs, defining the templates requires expertise and developers do not know how to design bug-triggering-prone templates, which could result in low efficiency and low effectiveness while testing. As complementary, our work could serve as background information when designing templates, boosting effectiveness and efficacy. Gligoric et al.~\cite{gligoric2013systematic} apply some refactorings in all applicable program elements in real-world projects without considering the characteristics of bug-triggering input programs, with our findings, one direction to improve their work is to mine bug-triggering input programs from real-world projects and prioritize the testing on those programs. The same situation also happens to \saferefactor-based tools~\cite{soares2009saferefactor,mongiovi2016scaling,soares2010making,soares2012automated,soares2011identifying,soares2009generating}, they test refactoring engines by generating unit tests to validate the behavior-preserving, however, refactoring should be validated on those error-prone input programs first. Existing studies~\cite{daniel2007automated,soares2009saferefactor,mongiovi2016scaling,soares2010making,soares2012automated,soares2011identifying,soares2009generating,gligoric2013systematic} to test the refactoring engines are unaware of the bug-triggering ability of the input programs, they are unguided and need to explore a large search space before triggering bugs. Our findings could serve as complementary knowledge to improve their effectiveness and efficacy. Alternatively, our findings could also be used to design new techniques for testing refactoring engines. For example, researchers could use our findings in Table~\ref{tab:characteristics} and the seed input programs from our dataset to construct prompts, and instruct the LLMs (e.g., ChatGPT) to generate more diverse input programs. In this way, it could potentially trigger more bugs in the refactoring engines. Based on our findings, more testing efforts should be paid to the input programs with language-specific features, considering \inputProgramJavaLanguageFeatureRatio{} of our studied bug reports are related to it, and its prevalence~\cite{dyer2014mining} in the OSS repositories.

\begin{tcolorbox}[left=0pt,right=0pt,top=0pt,bottom=0pt]
\textbf{Finding 9a:} Refactoring programs involving Java language features are more likely to trigger bugs in refactoring engines, they take up \inputProgramJavaLanguageFeatureRatio{} of our studied bug reports.

\end{tcolorbox}

Among all the language features, input programs containing Lambda expression (\inputProgramLambdaExpressionNumber{}/\inputProgramLambdaExpressionRatio{}), Java generics (\inputProgramGenericsNumber{}/\inputProgramGenericsRatio{}), and Enum (\inputProgramEnumNumber/\inputProgramEnumRatio{}) are the top three most bug-prone. While refactoring engines like \eclipse and \idea strive to provide robust support for modern Java language features, there can still be instances where refactorings involving generics, Lambda expressions, and Enum encounter difficulties due to the inherent complexity of these features and the limitations of refactoring engines. The main reasons are: (1) Lambda expressions and Java generics often involve type inference. It can be quite complex sometimes, especially in scenarios with nested generic types or where lambda expressions are involved. Refactoring tools need to accurately analyze and infer types to perform safe refactorings, and this complexity can lead to edge cases where the tools might not handle the refactoring correctly. For example, Figure~\ref{lst:lambdaexample}~\cite{lambdaexample} shows an example when extracting a method that contains a lambda expression, \idea fails to mark the parameter ``items'' as an input argument resulting in a compile error. This is because when performing control flow analysis, code fragments where the parent nodes are lambda types are not included, resulting in missing parameters. (2) Enums in Java are often used in ways that involve complex type hierarchies or switch statements. Refactorings involving enums need to handle these cases correctly to ensure that the behavior of the code remains consistent after the refactoring. Enum refactoring also involves handling references to enum constants and ensuring that they are updated correctly throughout the code base. (3) Refactoring tools typically analyze a limited scope of codebase to ensure reasonable performance. However, this limited scope can sometimes miss relevant dependencies or interactions that are crucial for correctly performing the refactoring, especially when dealing with features like lambda expressions and generics that involve complex type relationships.


\begin{figure}
\caption{Extract method in IntelliJ IDEA fails to mark ``items'' as input argument resulting in compile error}
\label{lst:lambdaexample}
{\centering
\begin{adjustbox}{width=0.5\textwidth}

\begin{lstlisting}[style=mystyle, escapechar=^]
default BitSet test(final List<T> items) {
    try {
        return ourPredicatesCache.get(Pair.create(this, items), () -> {
-      final BitSet result = new BitSet(items.size());
-       for (int i = 0; i < items.size(); i++) {
-         result.set(i, test(items.get(i)));
-       }
-       return result;
+       return getBitSet();
      });
    } catch (final ExecutionException e) {
      Logs.error(e);
      throw new RuntimeException(e);
    }
  }
+ @NotNull
+ default BitSet getBitSet() {
+   final BitSet result = new BitSet(items.size());
+   for (int i = 0; i < items.size(); i++) {
+   result.set(i, test(items.get(i)));
+   }
+   return result;
+ }
\end{lstlisting}\par
\end{adjustbox}\par
} 
\end{figure}

\begin{tcolorbox}[left=0pt,right=0pt,top=0pt,bottom=0pt]
\textbf{Finding 9b:} Lambda expression (\inputProgramLambdaExpressionNumber{}/\inputProgramLambdaExpressionRatio{}), Java generics (\inputProgramGenericsNumber{}/\inputProgramGenericsRatio{}), and Enum (\inputProgramEnumNumber/\inputProgramEnumRatio{}) are the top three language features triggering refactoring engine bugs due to the complexity of type inference, limited flow analysis, and complicated usage scenario.  
\end{tcolorbox}

Input programs with complex class relationships are the second most bug-triggering category. In this category, the inner class (\inputProgramInnerClassNumber{}/\inputProgramInnerClassRatio{}) and anonymous class (\inputProgramAnonymousClassNumber{}/\inputProgramAnonymousClassRatio{}) are more likely to induce bugs in refactoring engines. By analyzing the bug reports, input program, and usage scenario, we identified the following reasons: (1) Scope and visibility: inner classes and anonymous classes have unique scope and visibility rules compared to top-level classes. They can access both the final local variables of the enclosing block and members of their enclosing class, which might not be static. Managing these visibility and scope distinctions during refactorings, particularly when moving classes or changing accessibility, can complicate the refactoring logic. As shown in Figure~\ref{lst:anonymousclass}~\cite{anonymousclassexample}, when changing the method signature for a method inside an anonymous class, an exception occurs. After analyzing the patch, we find that this is because when visiting the AST node for AnonymousClassDeclaration wrongly stopped, thus resulting in an exception for the refactoring. (2) Complex naming and referencing: anonymous classes, by their nature, lack a named class declaration which makes them tricky in refactoring scenarios that involve class renaming or moving. Refactoring engines must handle references that are not tied to a simple class name but rather to their creation context, which can be nested deeply within methods or other classes. (3) Lifecycle and initialization concerns: refactoring that changes initialization order, such as moving an inner class outside or converting it to a static nested class, can inadvertently alter when and how instances of the class are created. This can lead to subtle bugs if the initialization sequence or the dependencies on the enclosing instance are not carefully managed.


\begin{figure}
\caption{Exception occurs when applying ``Change Method Signature'' on a method of an anonymous class in Eclipse 4.2.1}
\label{lst:anonymousclass}
{\centering
\begin{adjustbox}{width=0.5\textwidth}

\begin{lstlisting}[style=mystyle, escapechar=^]
public class ChangeMethodSignatureBug {
    public ChangeMethodSignatureBug(Object obj) {
    }
    public void m() {
        new ChangeMethodSignatureBug(new Object() {
            public void a(Object par1, Object par2) {
            }
        });
    }
}
\end{lstlisting}\par
\end{adjustbox}\par
} 
\end{figure}

\begin{tcolorbox}[left=0pt,right=0pt,top=0pt,bottom=0pt]
\textbf{Finding 10:} Input programs having complex class relationships are more likely to result in refactoring engine failure. Among these, refactoring involving inner class (\inputProgramInnerClassNumber{}/\inputProgramInnerClassRatio{}) and anonymous class (\inputProgramAnonymousClassNumber{}/\inputProgramAnonymousClassRatio{}) are the top two most bug-prone.  
\end{tcolorbox}

Annotation-induced (\inputProgramAnnotationNumber{}/\inputProgramAnnotationRatio{}) refactoring engine bugs are the third most in our studied bug reports with input program. Annotation is a frequently used feature in Java, and previous research~\cite{zhang2024understanding} has discovered that it can lead to bugs in program analysis and optimization tools.
By analyzing our studied bug reports together with the discussions and fixes, we summarize three factors that affect the parser, interpreter, and code modification of refactoring tools: (1) Incomplete semantics. Refactoring aims to alter the structure of the code without changing its semantics. However, annotations can carry semantic meaning that affects the program behavior, and refactoring tools may not understand and respect these semantics to avoid altering the program behavior. Figure~\ref{lst:annotations}~\cite{annotationexample} gives an example where push-down method refactoring produces a compilation error because @Override is not removed. @Override 
represents the relationship among classes in Java that allows a subclass or child class to provide a specific implementation of a method that is already provided by one of its super-classes or parent classes, and refactoring engine developers ignore it when rewriting the AST for program transformation, resulting in a compilation error as the @Override annotation is mistakenly kept during the refactoring. (2) Complexity of annotation processing. Some runtime annotations (e.g., @Deprecated) may influence the behavior of the program when it is running, which is typically outside the direct manipulation scope of static refactoring tools. (3) Complex transformations for annotations. Annotations often are used in contexts that involve reflection or dynamic code transformations, which are inherently complex to refactor. Refactoring tools must analyze and modify such annotations intelligently, a process that is error-prone due to the dynamic and interconnected nature of these frameworks.


\begin{figure}
\caption{``Push Down'' refactoring does not remove @Override leading to compilation error in Eclipse 4.2.1}
\label{lst:annotations}
{\centering
\begin{adjustbox}{width=0.3\textwidth}

\begin{lstlisting}[style=mystyle, escapechar=^]
class PushDownBug1 {
    static abstract class C {
        abstract void m();
    }
    static final C F = new C() {
        @Override
        void m() {
        }
    };
}
\end{lstlisting}
\end{adjustbox}
} 
\end{figure}

\begin{tcolorbox}[left=0pt,right=0pt,top=0pt,bottom=0pt]
\textbf{Finding 11:} Annotation-induced (\inputProgramAnnotationNumber{}/\inputProgramAnnotationRatio{}) refactoring engine bugs are the third most in our studied bug reports with input program. The reasons include: (1) Lack of consideration of the semantics of annotations, (2) Complexity of annotation processing, which may be processed at various phrases: compile-time (e.g., @Override), runtime (e.g., @Deprecated), and deployment (e.g. @WebServlet in Java Servlet API), (3) Complex transformations for annotations. 
\end{tcolorbox}

There are some (\inputProgramCommentRelatedNumber{}/\inputProgramCommentRelatedRatio{}) bugs related to code comments. For example, not updating the corresponding code comments after renaming the method parameters, accidentally deleting the code comments when extracting local variables, and refactoring tools might deny pulling up a variable as a field just because there is a code comment following the variable. After analyzing the attached patches, we identified the following reasons: (1) Comment-code dependency: comments sometimes rely on specific code constructs, such as method names, variable names, or parameter names, to maintain their relevance and accuracy. When refactoring tools modify these code elements, they may inadvertently render associated comments outdated or incorrect. For example, renaming a method parameter without updating the accompanying comment describing its purpose can lead to confusion. (2) Semantic gaps: comments often provide additional context or explanations about the code, but they are not formally part of the code's syntax or semantics. Refactoring engines primarily operate on the code's structure and semantics defined by the programming language, making it challenging to bridge the gap between code changes and their corresponding comments. For example, when \eclipse performs ``Move Class to New File'' refactoring, some comments are deleted in both the new file and the old file~\cite{moveClassRemoveCommentExample}. Figure~\ref{lst:commentrelated} shows an example where a code comment is wrongly deleted by \eclipse 3.6 when performing inline local variable refactoring. As the \eclipse developer stated: ``the problem comes from the fact that this segment is deleted by inlining operation without taking care of the attached comment''~\cite{commentrelatedexample}. An existing study~\cite{sommerlad2008retaining} tried to retain comments while refactoring code, but it is still limited in supporting multi-languages and cannot handle complex situations (e.g., code and comment tracing~\cite{wen2019large,codeCommentTracingExample}, parsing issues for input programs with comments~\cite{codeCommentParsingExample}, and considering various comment positions~\cite{codeCommentPositionExample}).


\begin{figure}
\caption{Comment is wrongly deleted when performing inline local variable refactoring using Eclipse 3.6}
\label{lst:commentrelated}
{\centering
\begin{adjustbox}{width=0.3\textwidth}

\begin{lstlisting}[style=mystyle, escapechar=^]
public void testIt() {
-  // blah
-  String test = "";
-  System.out.println(test);
+  System.out.println("");
}
\end{lstlisting}
\end{adjustbox}
} 
\end{figure}

\begin{tcolorbox}[left=0pt,right=0pt,top=0pt,bottom=0pt]
\textbf{Finding 12:} Refactoring involving comments interleaved with code might cause refactoring to fail, this contributes to \inputProgramCommentRelatedNumber{}/\inputProgramCommentRelatedRatio{} of our studied bugs. After analyzing the bug reports and patches, we conclude the reasons as: (1) Comment-code dependency, and (2) Semantic gaps between code and comments. This finding calls for further studies on comment-aware auto-refactoring tools.  
\end{tcolorbox}

\section{Transferability study}
\label{sec:crossstudy}

In this section, we conduct a transferability study using the historical refactoring engine bugs. Our hypothesis is that \emph{refactoring engines like \eclipse, \idea, and \netbeans all support the same frequently used refactoring types. Thus, for the common refactorings, bugs in one refactoring engine might also exist in other refactoring engines.} So, we could potentially uncover bugs by cross-validating the historical bug reports for each refactoring engine. Specifically, the procedures of our transferability study are following:

\begin{enumerate}

\item We first identify the refactoring types that are supported across all three studied refactoring engines: \eclipse, \idea, and \netbeans. We refer to these as Common Refactoring Types (CRT). From this set, we select the top 10 most error-prone refactoring types based on our Finding 1. The decision to focus on the top 10 is motivated by their higher likelihood of triggering bugs, as well as the fact that manually validating all CRTs would be non-trivial and time-consuming.

\item Bug reports related to the top 10 CRTs are extracted from the \refactorbench dataset and filtered for cross-validation. These bug reports are categorized into three partitions which correspond to bug reports from \eclipse, \idea, and \netbeans, respectively. For each partition, we cross-validate the bug reports against the other two target refactoring engines. For instance, bug reports from \eclipse are validated in both \idea and \netbeans by analogy.

\item We obtain the latest versions of \eclipse (2024-03), \idea (2024.1.2), and \netbeans (Apache NetBeans IDE 22) for testing. For each bug report within a partition, we first check the bug-tracking systems of the target IDEs to avoid duplicate reports before manual validation. The search process involves looking for refactoring type keywords and symptoms. If any bug reports with similar input programs are found, we consider the bug already reported and exclude it from further validation.

\item If no duplicated bug has been reported, we proceed with manual reproduction using the provided input program, and refactoring procedures. If any of the previously mentioned symptoms in RQ2 manifest, it is classified as a bug in the target refactoring engine.

\end{enumerate}

\begin{figure}[h]
    \centering
    \vspace{-3pt}    \includegraphics[width=0.6\textwidth]{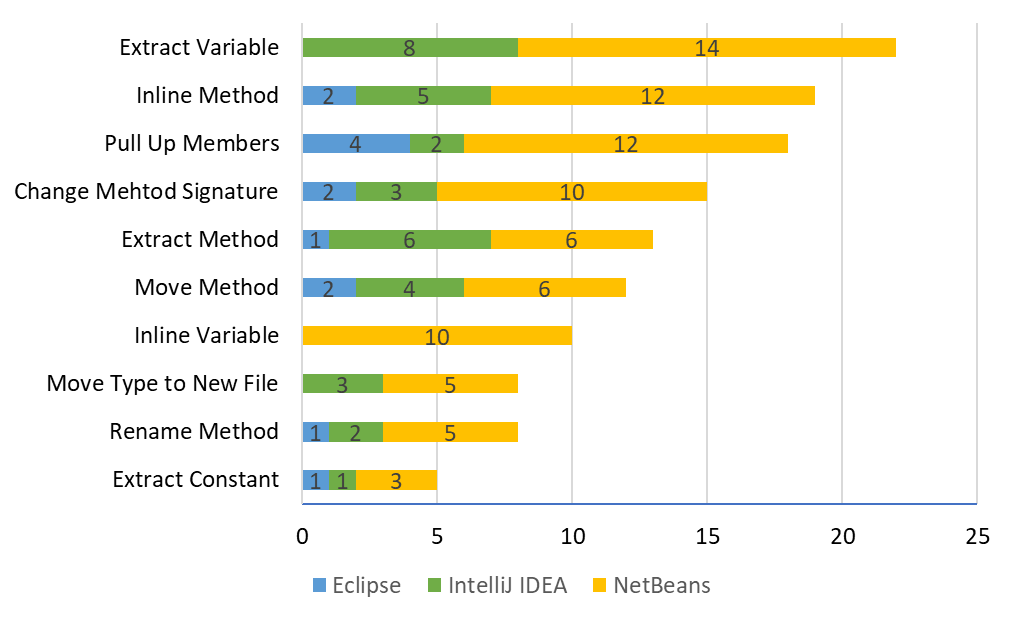}
    \vspace{-6pt}
     \caption{Bugs found by our transferability study across each target IDE}
    \label{newbugs}
\end{figure}

Figure~\ref{newbugs} illustrates the number of bugs identified in each target IDE during our transferability study for the top 10 refactoring types. For each target IDE, these bug reports represent new, previously unreported issues. In total, \foundIssueNumber{} bugs were uncovered through our study~\cite{foundBugs}. Specifically, 13 bugs were found in \eclipse, 34 in \idea, and 83 in \netbeans. The "Extract Variable" refactoring triggered the highest number of bugs (22), followed by the "Inline Method" (19 bugs) and "Pull Up Members" (18 bugs) refactorings. \netbeans accounts for the majority of bugs, with 83 issues, representing 63.8\% of the total bugs discovered. This suggests that \netbeans is less reliable compared to the others in handling refactorings. In contrast, \eclipse has the fewest bugs, with only 13 identified issues, making it the most robust refactoring tool among the three studied engines.

\begin{table*}
\centering
\caption{Submitted bugs revealed by our transferability study}
\label{tab:submittedbugreports}
\begin{adjustbox}{width=0.98\textwidth}
\begin{tabular}{ccccccc} 
\toprule
\textbf{ID} & \textbf{Source} & \textbf{Target} & \textbf{Issue No.} & \textbf{Title}                                                                                                                                                                                                                                                                                                                  & \textbf{Symptom}    & \textbf{Status}  \\ 
\hline
I-1         & Eclipse         & IDEA            & 354122             & Extract Method refactoring produces refactored program contains syntax error                                                                                                                                                                                                                                                    & Compile error       & Fixed            \\
I-2         & Eclipse         & IDEA            & 354116             & \begin{tabular}[c]{@{}c@{}}Refactoring erroneously qualifies calls inside the anonymous inheritor of the outer class\end{tabular}                                                                                                                                                                                 & Behavior change     & Fixed            \\
I-3         & Eclipse         & IDEA            & 354039             & Refactoring changes the behavior of my program                                                                                                                                                                                                                                                              & Behavior change     & Confirmed        \\
I-4         & Eclipse         & IDEA            & 354040             & Change Signature refactoring fail to perform the change as shown in the preview                                                                                                                                                                                                                                                 & Fail to refactoring & Confirmed        \\
I-5         & Eclipse         & IDEA            & 355272             & \begin{tabular}[c]{@{}c@{}}Pull Members Up refactoring for the classes using generic types \\producing uncompilable program\end{tabular}                                                                                                                                                                                        & Compile error       & Submitted        \\
I-6         & Eclipse         & IDEA            & 355271             & Move Inner Class to Upper Level refactoring fails for inner class using generic type                                                                                                                                                                                                                                            & Compile error       & Submitted        \\
I-7         & Eclipse         & IDEA            & 355273             & Move inner class to different package refactoring result in syntax error                                                                                                                                                                                                                                                        & Compile error       & Submitted        \\
I-8         & Eclipse         & IDEA            & 355200             & \begin{tabular}[c]{@{}c@{}}Move Inner Class to Upper Level refactoring produces uncompilable \\program for classes in different package\end{tabular}                                                                                                                                                                            & Compile error       & Submitted        \\
I-9         & Eclipse         & IDEA            & 354991             & \begin{tabular}[c]{@{}c@{}}Pull method up and make method abstract result in \\refactored program contains syntax error\end{tabular}                                                                                                                                                                                            & Compile error       & Submitted        \\
I-10        & Eclipse         & IDEA            & 355423             & "Introduce Variable" refactoring changes the behavior of the program                                                                                                                                                                                                                                                            & Behavior change     & Won't fix        \\
I-11        & Eclipse         & IDEA            & 355421             & \begin{tabular}[c]{@{}c@{}}\textcolor[rgb]{0.122,0.137,0.149}{Introduce Variable refactoring changes the semantic of the program because }\\\textcolor[rgb]{0.122,0.137,0.149}{IDEA does not identify the statements that }\\\textcolor[rgb]{0.122,0.137,0.149}{may change the value of the extracted expressions}\end{tabular} & Behavior change     & Submitted        \\
I-12        & Eclipse         & IDEA            & 354041             & Extract method refactoring issue about the parameter of the extracted method                                                                                                                                                                                                                                                    & Behavior change     & Won't fix        \\
I-13        & Eclipse         & IDEA            & 354042             & Type Migration refactoring produce refactored program contains syntax error                                                                                                                                                                                                                                                     & Compile error       & Shelved          \\
I-14        & Eclipse         & IDEA            & 353991             & Fail to rename class                                                                                                                                                                                                                                                                                                            & Fail to refactoring & Submitted        \\
E-1         & IDEA            & Eclipse         & 1529               & \begin{tabular}[c]{@{}c@{}}{[}Bug][Inline Method Refactoring] Inline the method which contians super \\keyword in a static method result in the refactored program has syntax error\end{tabular}                                                                                                                                & Compile error       & Fixed        \\
E-2         & IDEA            & Eclipse         & 1530               & \begin{tabular}[c]{@{}c@{}}{[}Bug][Move Static Memebers Refactoring] Move static members refactoring \\results in behavior change\end{tabular}                                                                                                                                                                                  & Behavior change     & Fixed        \\
E-3         & IDEA            & Eclipse         & 1531               & \begin{tabular}[c]{@{}c@{}}{[}Bug][Rename Refactoring] Avoid the perform of rename refactoring \\at the implicit enum elements (e.g. values() method)\end{tabular}                                                                                                                                                              & Compile error       & Fixed        \\
E-4         & IDEA            & Eclipse         & 1532               & {[}Bug][Pull Up Refactoring] Pull up refactoring produces uncompilable program                                                                                                                                                                                                                                                  & Compile error       & Fixed        \\
E-5         & IDEA            & Eclipse         & 1533               & \begin{tabular}[c]{@{}c@{}}{[}Bug][Pull Up Refactoring] Pull up method refactoring for \\method in the inner class fails\end{tabular}                                                                                                                                                                                           & Compile error       & Fixed        \\
N-1         & Eclipse         & NetBeans        & 7428               & Introducing method on "instanceof" caused compile error                                                                                                                                                                                                                                                                         & Compile error       & Submitted        \\
N-2         & Eclipse         & NetBeans        & 7427               & Fail to introduce method                                                                                                                                                                                                                                                                                                        & Fail to refactoring & Submitted        \\
\bottomrule
\end{tabular}
\end{adjustbox}
\begin{tablenotes}
\footnotesize
\item{
The issues of \idea, \eclipse, and \netbeans can be found at https://youtrack.jetbrains.com/issue/IDEA-XXX, https://github.com/eclipse-jdt/eclipse.jdt.ui/issues/XXX, and https://github.com/apache/netbeans/issues/XXX, where ``XXX'' can be replaced with the concrete numbers in \textbf{Issue No.}. ``Shelved'' means the bug is confirmed but it is not included in the foreseen product plans.
}
\end{tablenotes}
\end{table*}

Table~\ref{tab:submittedbugreports} lists the \submittedIssueNumber{} bugs we submitted by the time of manuscript submission. We are currently preparing for the submission of the rest of the bugs. The ``Source'' column indicates the origin of the historical bug report, while the ``Title'' column specifies the refactoring tool being tested. The ``Issue No.'' and ``Title'' columns provide detailed information on each submitted issue, and the ``Symptom'' column summarizes the symptom associated with each bug. The final column details the current status of each submitted issue. By the time of this paper's submission, \submittedIssueFixedNumber{} of the issues had been fixed, and \submittedIssueConfirmedNumber{} had been confirmed.

\begin{figure}
\caption{``Introduce Variable'' refactoring changes the behavior of the input program in \idea 2024.1.2}
\label{lst:ideaIncorrectIntroduceVariable}
{\centering
\begin{adjustbox}{width=0.4\textwidth}

\begin{lstlisting}[style=mystyle, escapechar=^]
public class A {
    static String[] arr = {"1", "2", "3", "4", "5"};
    static int i = 0;
    private static String foo(String str) {
        return str + ":" + arr[i++];
    }
    public static void main(String[] args) {
-        System.out.println(foo("hello"));
-        System.out.println(foo("hello"));
-        System.out.println(foo("hello"));
-        System.out.println(foo("hello"));
+        String hello = foo("hello");
+        System.out.println(hello);
+        System.out.println(hello);
+        System.out.println(hello);
+        System.out.println(hello);

    }
}
\end{lstlisting}\par
\end{adjustbox}\par
} 
\end{figure}

Figure~\ref{lst:ideaIncorrectIntroduceVariable} illustrates the input program and the refactored result produced by \idea 2024.1.2 for issue I-10 in Table~\ref{tab:submittedbugreports}. Before refactoring, the program's output was: ``hello:1, hello:2, hello:3, hello:4''. However, the output of the refactored program became: ``hello:1, hello:1, hello:1, hello:1.'' This occurred because each invocation of the ``foo'' method updates the state of ``i'', producing different outputs for each call. However, \idea extracted ``foo("hello")'' as a variable and replaced each invocation with the extracted variable, disregarding the side effects of the extracted expression. As a result, the output was incorrect: ``hello:1, hello:1, hello:1, hello:1'', as the ``foo'' method is only invoked once in the refactored program. The original bug was reported in \eclipse~\cite{eclipseExtractVariableExample}, and the developers addressed the issue by implementing side effect analysis during variable extraction. Although we identified the same issue in the latest version of \idea and reported it to its developers, they classified it as a usability problem and unwilling to fix it by explaining that ``We don’t guarantee that there will be no semantics change when `all occurrences' is selected, and it is the user's responsibility to decide whether it is safe or not''~\cite{willnotfixissueexample}. A similar issue was observed in I-12 in Table 6. This suggests that while different IDEs support the same refactoring types, their handling of complex operations, such as side effect analysis, can vary slightly. Nevertheless, during our interactions with refactoring engine developers, they largely agreed with our reported bugs, resulting in \submittedIssueFixedNumber{} fixes and \submittedIssueConfirmedNumber{} confirmations. One issue (I-13) was confirmed by the \idea developers but was not included in their immediate product plans, leading them to assign it the status of ``Shelved''.




\section{Threats to Validity}
We identify the following threats to the validity of this paper:

\noindent \textbf{Internal.} The internal threat to validity mainly lies in our manual classification and labeling of refactoring engine bugs, which may have subjective bias or errors. To reduce this threat, we referred to the previous studies~\cite{sun2016toward,shen2021comprehensive,yang2011finding}, and then adopted an open-coding scheme to derive the taxonomies to fit refactoring engine bugs. During the labeling process, the first two authors independently labeled bugs, any disagreement was discussed at a meeting until a consensus was reached. 

\noindent\textbf{External.} The external threat to validity mainly lies in the datasets used in our study. To reduce this threat, we systematically collected refactoring engine bugs as presented in Section 3. To guarantee the generalizability of our study, we used three popular refactoring engines as subjects and studied \numberofBugsWithInputProgram{} bugs in total by balancing the effort of manual analysis and the study scale.
\section{Implication}

Based on our study and analysis, we discuss the implications for the refactoring engine developers and researchers.

\noindent \textbf{Implication for Developers.}
Based on Finding 1, ``Extract'', ``Inline'', and ``Move'' refactorings are the most error-prone. Refactoring engine developers should focus on improving the reliability of these types, expanding test coverage, and considering more diverse input programs. Current testing tools only cover about one-third of the refactoring types, leaving gaps that could be targeted for improvement. Finding 2 emphasizes that the most common bug symptoms are compile errors, crashes, and behavior changes. Developers should enhance test oracles, particularly for detecting behavior changes, which pose significant challenges. Furthermore, accounting for side effects such as warnings and refactoring availability (Finding 3) should also be integrated into error-checking processes to ensure consistency. Finding 4 highlights the prominence of incorrect transformations as a root cause of bugs. Developers should work towards diverse input programs and leverage data-driven automation to minimize these issues. Additionally, automating the inference of preconditions (Finding 5) and enhancing flow analysis and type resolution mechanisms (Finding 6 and Finding 7) would help mitigate errors. Given that refactoring involving modern Java language features triggers a significant number of bugs (Finding 9a), refactoring engines should improve their handling of language-specific features like lambda expression, generic, and Enum (Finding 9b). These features present complexities in type inference and flow analysis, which should be better addressed in future engine iterations. Finding 10 shows that complex class structures, such as inner and anonymous classes, are prone to triggering bugs. Similarly, annotations (Finding 11) add complexity due to their semantic meaning and wide-ranging effects. Refactoring engines should be enhanced to handle these constructs more effectively, particularly by improving transformations that respect class and annotation intricacies. Finding 12 emphasizes the challenges posed by comments interleaved with code, where dependencies between code and comments can lead to semantic gaps. Developers should explore new methods to make refactoring tools more comment-aware, ensuring that the intended meaning of both code and comments is preserved during refactoring.

\noindent \textbf{Implication for Researchers.}
Based on Finding 1, future research should explore improved testing techniques that focus on the most error-prone refactoring. These techniques should go beyond existing tools to cover a wider range of refactoring types, leveraging the findings as a basis for designing more effective tests. Finding 2 points out the challenges in detecting behavior change bugs. Researchers should focus on how to effectively and efficiently identify behavior changes. Besides, incorporating side effects, such as refactoring availability or warning messages, into bug detection strategies could provide a broader perspective on refactoring correctness. As mentioned in Findings 4 and 5, automated, data-driven approaches are a promising direction for improving refactoring engine testing. Researchers could investigate large language models (LLMs) to generate diverse input programs. Finding 8 reveals that most of the refactoring engine bugs could be triggered by the default configuration, thus relieving researchers from testing configuration-related refactoring bugs. Finding 9a and Finding 9b highlight that language features, particularly lambda expressions and generics, significantly contribute to refactoring bugs. Researchers should incorporate new language features into their testing techniques to ensure a better coverage of modern programming language features. Finding 10 and 11 suggest that complex class relationships and annotations are common triggers for refactoring bugs. Future research could explore how refactoring engines can be improved to handle them better. Based on Finding 12, there is a gap in research regarding comment-aware refactoring. Researchers should investigate new methodologies for creating tools that can bridge the semantic gap between code and comments, enabling more robust refactoring that considers code-comment dependencies.
\section{Related Work}

\noindent \textbf{Detecting Refactoring Engine Bugs.}
There are several techniques for testing refactoring engine bugs in the literature~\cite{daniel2007automated,soares2009saferefactor,mongiovi2016scaling,soares2010making,soares2012automated,soares2011identifying,soares2009generating,gligoric2013systematic}. For example, Daniel et al.~\cite{daniel2007automated} developed ASTGen, a widely-used Java test program generation tool, to test refactoring engines. Soares et al.~\cite{soares2009saferefactor} proposed SAFEREFACTOR, which relies on random tests to make sure the refactorings are behavior-preserving. Soares et al.~\cite{mongiovi2016scaling,soares2010making,soares2012automated,soares2011identifying} proposed testing refactoring engines based on SAFEREFACTOR. Meanwhile, Gligoric et al.~\cite{gligoric2013systematic} tested Eclipse refactoring engines for Java (JDT) and C (CDT) in an end-to-end approach on real software projects. Given a set of projects, they apply refactorings in many places and collect failures where the refactoring engine throws an exception or produces refactored code that does not compile. Eric et al.~\cite{lacker2021statistical} analyzed the existing period, fixing ratio, and duplication percentage of refactoring bugs of \jdt. To validate the effectiveness of the refactoring detection tools (i.e., RefactoringMiner~\cite{tsantalis2020refactoringminer} and RefDiff~\cite{silva2020refdiff}), Leandro et al.~\cite{leandro2022technique} first apply refactorings using
a popular IDE (i.e., \eclipse) and then run those two refactoring detection tools to
check whether they can detect the transformations performed by the IDE. Different from existing research in refactoring bugs detection, our work focuses on studying the characteristics of refactoring engine bugs to facilitate testing and debugging. Current approaches may be ineffective as (1) they randomly generate test programs or apply refactoring without any guidance, and (2) lacking a general and in-depth understanding
of bugs in the refactoring engines. Therefore, as the first empirical study on refactoring engine bugs, we believe that our work helps design effective testing and debugging methods based on our findings.

\noindent\textbf{Empirical Studies on Bugs.} Researchers have conducted various empirical studies to understand the characteristics of bugs, including compilers~\cite{shen2021comprehensive, sun2016toward,zhong2022enriching,xu2023silent}, Android~\cite{xiong2023empirical,bhattacharya2013empirical,linares2017empirical}, static analyzers~\cite{zhang2023statfier,zhang2024understanding}, etc. For example, Sun et al.~\cite{sun2016toward} conducted an empirical study to analyze the duration, priority, fixes, test cases, and locations of GCC and LLVM bugs. Zhong et al.~\cite{zhong2022enriching} study the historical compiler bug reports and enrich compiler testing with real programs from bug reports. Xiong et al.~\cite{xiong2023empirical} study the functional bugs in Android apps. Zhang et al.~\cite{zhang2024understanding} study the annotation-induced faults for static analyzers like PMD. Different from the previous studies, we target refactoring engine bugs. To the best of our knowledge, we are the first to characterize refactoring engine bugs focusing on various perspectives (symptoms, root causes, triggering conditions). Our study helps in gaining better understanding of the refactoring engine bugs, which can potentially benefit developers of refactoring engines and researchers of refactoring engines bugs detection. Based on our findings and dataset, our transferability study has revealed more than 100 new bugs in the latest version of refactoring tools.

\section{Conclusion}
As integral components of modern Integrated Development Environments (IDEs), refactoring engines can alleviate the burden of development and optimization by assisting developers in restructuring their code. However, it is inevitable for these engines to have bugs, like traditional software systems. Their bugs could be propagated to the system refactored by them and produce unexpected, even dangerous behaviors during real-world usage. Therefore, it is necessary to guarantee the quality of refactoring engines. In this paper, we conducted the first comprehensive study to understand refactoring engine bugs to promote the design of effective bug detection and debugging techniques. Specifically, we manually studied \numberofBugsWithInputProgram{} bugs from three popular refactoring engines, identified several root causes, bug symptoms, and input program characteristics, and obtained \numberofFindings{} major findings. Based on these findings, we provide a set of guidelines for refactoring engine bug detection and debugging. Our transferability study reveals \foundIssueNumber{} new and unique bugs in the latest version of refactoring tools. Among the \submittedIssueNumber{} bugs we submitted, \totalconfirmedIssueNumber{} bugs are confirmed by their developers, and \submittedIssueFixedNumber{} of them have already been fixed.

\noindent\textbf{Data Availability.}
The data is available at ~\cite{ourrepourl}.

\section*{Acknowledgments}
This work is
supported by the Natural Sciences and Engineering Research Council of Canada (NSERC) Discovery Grants.

\bibliographystyle{ACM-Reference-Format}
\bibliography{refs}

\end{document}